\documentclass[10pt]{bmc_article}

\usepackage{cite} % Make references as [1-4], not [1,2,3,4]
\usepackage{url}  % Formatting web addresses
\usepackage{ifthen}  % Conditional
\usepackage{multicol}   %Columns
\usepackage[utf8]{inputenc} %unicode support

\usepackage[cmex10]{amsmath}
\usepackage{amssymb}
\usepackage{dsfont}
\usepackage{array}
\usepackage[ruled,vlined]{algorithm2e}

\usepackage{multirow}
\usepackage{graphicx}
\usepackage{amsthm}

\newtheorem{theorem}{Theorem}
\newtheorem{corollary}{Corollary}

\newboolean{publ}

\newenvironment{bmcformat}{\baselineskip20pt\sloppy\setboolean{publ}{false}}{\baselineskip20pt\sloppy}

\begin{document}
\begin{bmcformat}


\title{Achieving Low-Complexity Maximum-Likelihood Detection for the 3D MIMO Code}

%%%%%%%%%%%%%%%%%%%%%%%%%%%%%%%%%%%%%%%%%%%%%%
%%                                          %%
%% Enter the authors here                   %%
%%                                          %%
%% Ensure \and is entered between all but   %%
%% the last two authors. This will be       %%
%% replaced by a comma in the final article %%
%%                                          %%
%% Ensure there are no trailing spaces at   %%
%% the ends of the lines                    %%     	
%%                                          %%
%%%%%%%%%%%%%%%%%%%%%%%%%%%%%%%%%%%%%%%%%%%%%%

\author{Ming~Liu$^1$%
         \email{Ming Liu - ming.liu@insa-rennes.fr}
       ,
         Matthieu~Crussi\`ere\correspondingauthor$^1$%
         \email{Matthieu~Crussi\`ere\correspondingauthor - matthieu.crussiere@insa-rennes.fr}%
       ,
        Maryline~H\'elard$^1$%
        \email{Maryline~H\'elard - maryline.helard@insa-rennes.fr}%
       and
        Jean-Fran\c{c}ois~H\'elard$^1$%
        \email{Jean-Fran\c{c}ois~H\'elard - jean-francois.helard@insa-rennes.fr}%
      }

%%%%%%%%%%%%%%%%%%%%%%%%%%%%%%%%%%%%%%%%%%%%%%
%%                                          %%
%% Enter the authors' addresses here        %%
%%                                          %%
%%%%%%%%%%%%%%%%%%%%%%%%%%%%%%%%%%%%%%%%%%%%%%

\address{%
    \iid(1)Universit\'e Europ\'eenne de Bretagne (UEB), INSA, IETR, UMR 6164, 20 avenue des Buttes de Co\"esmes, F-35708 Rennes, France
}%

\maketitle

%%%%%%%%%%%%%%%%%%%%%%%%%%%%%%%%%%%%%%%%%%%%%%
%%                                          %%
%% The Abstract begins here                 %%
%%                                          %%
%% Please refer to the Instructions for     %%
%% authors on http://www.biomedcentral.com  %%
%% and include the section headings         %%
%% accordingly for your article type.       %%
%%                                          %%
%%%%%%%%%%%%%%%%%%%%%%%%%%%%%%%%%%%%%%%%%%%%%%

\begin{abstract}
        % Do not use inserted blank lines (ie \\) until main body of text.
The 3D MIMO code is a robust and efficient space-time block code (STBC) for the distributed MIMO broadcasting but suffers from high maximum-likelihood (ML) decoding complexity.
In this paper, we first analyze some  properties of the 3D MIMO code to show that the 3D MIMO code is fast-decodable.
It is proved that the ML decoding performance can be achieved with a complexity of $O(M^{4.5})$ instead of $O(M^8)$ in quasi static channel with $M$-ary square QAM modulations.
Consequently, we propose a simplified ML decoder exploiting the unique properties of 3D MIMO code.
Simulation results show that the proposed simplified ML decoder can achieve much lower processing time latency compared to the classical sphere decoder with Schnorr-Euchner enumeration.
\end{abstract}

\ifthenelse{\boolean{publ}}{\begin{multicols}{2}}{}

\section*{Keywords}
MIMO, space-time codes, maximum likelihood decoding, computational complexity.

%%%%%%%%%%%%%%%%%%%%%%%%%%%%%%%%%%%%%%%%%%%%%%
%%                                          %%
%% The Main Body begins here                %%
%%                                          %%
%% Please refer to the instructions for     %%
%% authors on:                              %%
%% http://www.biomedcentral.com/info/authors%%
%% and include the section headings         %%
%% accordingly for your article type.       %%
%%                                          %%
%% See the Results and Discussion section   %%
%% for details on how to create sub-sections%%
%%                                          %%
%% use \cite{...} to cite references        %%
%%  \cite{koon} and                         %%
%%  \cite{oreg,khar,zvai,xjon,schn,pond}    %%
%%  \nocite{smith,marg,hunn,advi,koha,mouse}%%
%%                                          %%
%%%%%%%%%%%%%%%%%%%%%%%%%%%%%%%%%%%%%%%%%%%%%%

%%%%%%%%%%%%%%%%
%% Background %%
%%
\section{Introduction}
Multiple-input multiple-output (MIMO) is a promising technique that can bring significant improvements to the wireless communication systems. %to break the barriers of Shannon's limit.
In combination with space-time block code (STBC), it provides higher spectrum efficiency with better communication reliability~\cite{tarokh1998space}.
In the last decades, MIMO has been widely employed in the latest wireless communication standards such as IEEE 802.11n, 3GPP Long Term Evolution (LTE), WiMAX and Digital Video Broadcasting--Next Generation Handheld (DVB-NGH) etc.
It is also seen as the key technology for the future digital TV terrestrial broadcasting standards~\cite{DVB-T2-MIMO}.

A so-called space-time-space (3D) MIMO code~\cite{nasser20083d} was proposed for the future TV broadcasting systems in which the services are delivered by the MIMO transmission in a single frequency network (SFN).
Specifically, it is proposed for a distributed MIMO broadcasting scenario where TV programs are transmitted by two geographically separated transmission sites, each site equipping two transmit antennas.
On the other hand, each receiver has two receive antennas, forming a $4\times2$ MIMO transmission.
The 3D MIMO code has been shown to be robust and efficient in the distributed MIMO broadcasting scenarios where there exists strong received signal power imbalances~\cite{liu2013distributed}.
Hence, it is a promising candidate for the MIMO profile of future broadcasting standards.
However, the 3D MIMO code suffers from a high computational complexity when the maximum-likelihood (ML) decoding is adopted.
The decoding complexity is as high as $O(M^8)$ when $M$-QAM constellation is used.
Up to now, no study on the decoding complexity reduction for the 3D MIMO code has been carried out in the literature.

Recently a lot of efforts have been made in the STBC design to obtain both high code rate and low decoding complexity~\cite{sharma2004full,dao2008four,biglieri2009fast,srinath2009low,sinnokrot2008embedded,ren2010fast,ismail2013new}.
The decoding complexity reduction is commonly achieved by exploiting the orthogonality embedded in the STBC codeword.
When there exists group-wise orthogonality in the codeword, the joint detection of many information symbols is converted into independent, group-wise detections~\cite{dao2008four,ren2010fast}, yielding low decoding complexity.
For other cases such as DjABBA code~\cite{hotinen2003multiantenna}, Biglieri-Hong-Viterbo (BHV) code~\cite{biglieri2009fast}, Srinath-Rajan code~\cite{srinath2009low} and Ismail-Fiorina-Sari (IFS) code~\cite{ismail2013new} in which the orthogonality only exists in a part of information symbols, some symbols can be detected in a group-wise manner once we condition them with respect to other symbols.
The ML solutions can be obtained with a lower complexity compared with the ML detector.
In other words, their decoding complexity is less than $O(M^{\kappa})$ where $\kappa$ is the number of information symbols in a codeword.
Such kind of STBCs are referred to as fast decodable STBCs~\cite{biglieri2009fast}.
However, most of fast-decodable STBCs are not optimized for distributed MIMO broadcasting scenarios and they are not robust under the received signal power imbalance conditions~\cite{liu2013distributed}.

A partial interference cancellation (PIC) group decoding scheme has been presented aiming at reducing the decoding complexity of the STBCs containing group-wise orthogonalities in the codewords~\cite{dai2005efficient,guo2009onfull}.
A number of STBCs that are optimized for this decoding scheme have also been proposed~\cite{guo2009onfull,zhang2012two}.
This scheme actually uses a linear equalization to convert the joint detection of a large number of symbols to several groups of ML decodings for few symbols.
However, the overall performance of this decoding scheme cannot achieve the ML optimality.

Some alternatives with reduced decoding complexity have been presented for the distributed MIMO broadcasting.
Polonen \emph{et al.} described a STBC with less decoding complexity based on orthogonal basis~\cite{polonen2011reduced}. However, such a code does not achieve full-diversity or full-rate for $4\times2$ MIMO transmissions and therefore performs worse than 3D MIMO code.
A ``punctured version'' of 3D MIMO code that is full-rate for $4\times2$ MIMO transmissions with low decoding complexity has also been proposed~\cite{liu2012distributed}.
However, it does not achieve full-diversity and is hence less robust in harsh channel conditions.

In this paper, we propose a reduced-complexity ML decoder for the 3D MIMO code which exploits the embedded orthogonality in the codeword.
The main contributions are:
\begin{itemize}
  \item We propose to modify the original 3D MIMO codeword through some permutations of information symbols which leads to an ML decoding algorithm with reduced complexity without affecting all desirable properties of the 3D MIMO code.
  \item We prove that the 3D MIMO code is fast decodable. Moreover, we show that the worst case decoding complexity is $O(M^{4.5})$ for $M$-ary square QAM modulations which is the least among all square full-rate  STBCs for $4\times2$ MIMO transmission.
  \item Based on the unique properties of the new form of 3D MIMO codeword, we propose a novel implementation of the simplified decoder that achieves a lower average complexity in terms of time latency without losing the ML optimality. The proposed implementation is also applicable for other fast decodable STBCs.
\end{itemize}

The remainder of the paper is organized as follows.
Some fundamentals of the MIMO detection are presented in Section~\ref{sec:sys_mdl}.
In Section~\ref{sec:3DMIMO}, the 3D MIMO code is first recalled.
Consequently, a modification of the codeword is proposed to facilitate the decoding process.
Three important properties of the new codeword are also revealed.
With this knowledge, in Section~\ref{sec:MLdec}, the ML decoder with a worst case decoding complexity of $O(M^{4.5})$ is derived.
Then in Section~\ref{sec:ImpMLdec}, a new implementation of the reduced-complexity ML decoder is described.
Section~\ref{sec:simu} presents the symbol error and complexity performance of the new decoder.
Conclusions are drawn in Section~\ref{sec:conclusion}.

\emph{Notations:} Vectors and matrices are written in boldface letters. Superscript $\mathbf{X}^T$ represents transposition of matrix $\mathbf{X}$.  $x^R$ and $x^I$ denote the real and imaginary parts of a complex number $x$, respectively.
The operator $(\check{\cdot})$ performs the complex-real conversion from $\mathbb{C}$ to $\mathbb{R}^{2\times2}$:
\begin{equation}\label{eq:check_function}
    \check{x} \triangleq \left [\begin{array}{*{20}c}
      x^R & -x^I \\
      x^I & x^R \\
    \end{array}\right ].
\end{equation}
When $(\check{\cdot})$ operator is applied to a matrix $\mathbf{X}\in \mathbb{C}^{m\times n}$, the operation in (\ref{eq:check_function}) is performed for all elements $x_{j,k}$'s in the matrix, i.e. the $(j,k)$th $2\times2$ submatrix of $\check{\mathbf{X}}$ is $\check{x}_{j,k}$.
For a complex vector $\mathbf{x}=[x_1,x_2,\ldots, x_n]^T\in\mathbb{C}^n$, the operator $(\tilde{\cdot})$ separates the real and imaginary parts of the given vector, i.e. $\widetilde{\mathbf{x}}\triangleq[x_1^R,x_1^I,\ldots, x_n^R,x_n^I]^T$.
For a matrix $\mathbf{X}=[\mathbf{x}_1,\mathbf{x}_2,\ldots, \mathbf{x}_n]$ where $\mathbf{x}_j$ is the $j$th column of $\mathbf{X}$, the operator  $vec(\mathbf{X})$ stacks the columns of $\mathbf{X}$ to form one column vector, i.e. $vec(\mathbf{X})\triangleq[\mathbf{x}_1^T,\mathbf{x}_2^T,\ldots, \mathbf{x}_n^T]^T$.
$\widetilde{vec(\mathbf{X})}$ denotes vectorizing matrix $\mathbf{X}$ followed by the real/imaginary part separation.
The inner product of two real-valued vectors $\mathbf{x}$ and $\mathbf{y}$ is denoted by $\langle \mathbf{x},\mathbf{y}\rangle=\mathbf{x}^T\mathbf{y}$.
The $n\times n$ identity matrix is denoted by $\mathbf{I}_{n}$.
The operator $\otimes$ denotes the Kronecker product.
Finally, $i$ represents $\sqrt{-1}$.

\section{System Model}
\label{sec:sys_mdl}
\subsection{MIMO system model}
We consider a MIMO transmission with $N_t$ transmit and $N_r$ receive antennas over flat-fading channel. The received signal $\mathbf{Y}\in \mathbb{C}^{N_r\times T}$ is:
\begin{equation}\label{eq:rec_sig}
    \mathbf{Y}=\mathbf{H}\mathbf{X}+\mathbf{W},
\end{equation}
where $\mathbf{X}\in \mathbb{C}^{N_t\times T}$ is the STBC codeword matrix which is transmitted over $T$ channel uses; $\mathbf{W}\in \mathbb{C}^{N_r\times T}$ is a complex-valued additive white Gaussian noise (AWGN) component; $\mathbf{H}\in \mathbb{C}^{N_r\times N_t}$ is the channel matrix whose $(j,k)$th element $h_{j,k}$ denotes the channel coefficient of the link between the $k$th transmit antenna and the $j$th receive antenna.
The channel is assumed to be quasi-static. That is, the channel coefficients keep constant over the duration of one STBC codeword, but change from one codeword to another.
Moreover, $h_{j,k}$'s are assumed to be independent from each other.

For linear STBCs, the codeword matrix $\mathbf{X}$ can be obtained through a linear operation~\cite{biglieri2009fast}:
\begin{equation}\label{eq:ST_coding_by_generator_matrix}
    \widetilde{vec(\mathbf{X})} = \mathbf{G}\widetilde{\mathbf{s}},
\end{equation}
where $\mathbf{s}=[s_1, s_2,\ldots,s_{\kappa}]^T$ is the vector containing $\kappa$ independent information symbols.
The code rate of  STBC is $\kappa/T$ information symbols per channel use.
The generator matrix $\mathbf{G}\in \mathbb{R}^{2N_tT\times2\kappa}$ is obtained:
\begin{equation}\label{eq:generator_matrix}
    \mathbf{G}\triangleq[\widetilde{vec(\mathcal{A}_1)}, \widetilde{vec(\mathcal{B}_1)},\ldots,\widetilde{vec(\mathcal{B}_\kappa)}],
\end{equation}
where $\mathcal{A}_j\in \mathbb{C}^{N_t\times T}$ and $\mathcal{B}_j\in \mathbb{C}^{N_t\times T}$ are the complex weight matrices representing the contribution of the real and imaginary parts of the $j$th information symbol $s_j$ in the final codeword matrix.

Separating the real and imaginary parts of the transmitted and received signals, and stacking the columns of the codeword, the received MIMO signal (\ref{eq:rec_sig}) can be expressed in an equivalent real-valued form:
\begin{equation}\label{eq:rec_sig_real_eq}
    \widetilde{\mathbf{y}}=\mathbf{H}_{eq}\widetilde{\mathbf{s}}+\widetilde{\mathbf{w}},
\end{equation}
where $\widetilde{\mathbf{y}} = \widetilde{vec(\mathbf{Y})}$, $\widetilde{\mathbf{w}} = \widetilde{vec(\mathbf{W})}$ and $\mathbf{H}_{eq}\in \mathbb{R}^{2N_rT\times2\kappa}$is the equivalent channel matrix and is obtained by:
\begin{equation}\label{eq:Heq}
  \mathbf{H}_{eq}=(\mathbf{I}_{T}\otimes\check{\mathbf{H}})\mathbf{G}.
\end{equation}
Note that the real-valued expression of the signal can be obtained from the complex-valued form via a linear transform. Hence, we will jointly use both real- and complex-valued forms in the sequel.

\subsection{ML decoding of MIMO signals}
Once the channel $\mathbf{H}_{eq}$ is known by the receiver\footnote{We assume that the receiver has perfect knowledge of the channel in our work. In practice, the channel coefficients should be estimated using some channel estimation techniques. }, the information symbols can be retrieved from the received signal $\widetilde{\mathbf{y}}$ in (\ref{eq:rec_sig_real_eq}).
The maximum-likelihood (ML) solution of the transmitted signal is the combination of information symbols $\widetilde{\mathbf{s}}=[s_1, s_2,\ldots,s_\kappa]$ that minimizes the Euclidian distance between the channel distorted information signal $\mathbf{H}_{eq}\widetilde{\mathbf{s}}$ and received signal $\widetilde{\mathbf{y}}$, namely:
\begin{equation}\label{eq:ML_detection}
    \hat{\mathbf{s}}^{\mathrm{ML}}=\arg\min_{\mathbf{s}\in\boldsymbol\Theta^\kappa}\|\widetilde{\mathbf{y}}-\mathbf{H}_{eq}\widetilde{\mathbf{s}} \|^2,
\end{equation}
where $\boldsymbol\Theta$ is the set of the constellation symbols.
(\ref{eq:ML_detection}) indicates that the ML solution is found by \emph{jointly} determining $\kappa$ independent information symbols.
In other words, when the modulation of these symbols is $M$-QAM, the ML decoding should exhaustively check all $M^\kappa$ combinations. The search complexity grows dramatically with higher modulation order or larger number of information symbols in one codeword.
Hence, the ML decoding is computationally demanding.

\subsection{Fast ML decoding of MIMO signals}

More efficient STBC decoding is achieved with the help of orthogonal-triangular (QR) decomposition~\cite{agrell2002closest,biglieri2009fast}.
The QR decomposition of the equivalent channel matrix $\mathbf{H}_{eq}$ yields $\mathbf{H}_{eq}=\mathbf{Q}\mathbf{R}$, where
$\mathbf{Q}\in \mathbb{R}^{2N_rT\times2\kappa}$ is a unitary matrix, and $\mathbf{R}\in \mathbb{R}^{2\kappa\times2\kappa}$ is an upper triangular matrix.
The definitions of the ``classical Gram-Schmidt algorithm'' based QR decomposition can be found in the Appendices.
Note that other numerically stable QR decomposition algorithms can also be used without affecting the properties of the 3D MIMO code as  well as the resulting low-complexity decoding methods.
Instead of solving (\ref{eq:ML_detection}), the ML solution can be alternatively found by:
\begin{equation}\label{eq:SD_detection_metric}
    \hat{\mathbf{s}}^{\mathrm{ML}}=\arg\min_{\mathbf{s}\in\boldsymbol\Theta^\kappa\cap\mathcal{S}}\|\widetilde{\mathbf{z}}-\mathbf{R}\widetilde{\mathbf{s}}\|^2,
\end{equation}
where $\widetilde{\mathbf{z}}=\mathbf{Q}^{T}\widetilde{\mathbf{y}}\in\mathbb{R}^{2\kappa}$ is a linear transformation of received signal; $\mathcal{S}$ is a hypersphere centered on the received signal.  Only the codewords inside the hypersphere are checked during the search in order to reduce the search complexity.
The size of the hypersphere is represented by its radius.
The decoding process is turned into a bounded search over a $\kappa$-level tree with complex-valued nodes.
Hence, the worst case decoding complexity is $O(M^{\kappa})$.

Moreover, according to the property of QR decomposition, some information symbols can be decoded independently from the others if some elements of $\mathbf{R}$ are equal to zero.
It suggests that the joint search in a high dimension is converted into a bunch of parallel, independent searches in low dimensions.
This results in a significant reduction of the worst-case decoding complexity~\cite{biglieri2009fast,srinath2009low,sinnokrot2010fast}.

%\section*{Section title}
%\subsection*{Sub-heading for section}
%\subsubsection*{Sub-sub heading for section}
%\subsubsection*{Sub-sub-sub heading for section}

%\bigskip

\section{3D MIMO Code}
\label{sec:3DMIMO}
The 3D MIMO code~\cite{nasser20083d} possesses a better robustness against receive signal power imbalances in the distributed MIMO broadcasting scenarios but suffers from high decoding complexity~\cite{liu2013distributed}.
In this section, we propose a new 3D MIMO codeword that enables low sphere decoding complexity via exchanging the positions of information symbols in the original 3D MIMO codeword.
The basic idea behind this modification comes from the facts that the orthogonality embedded in the information symbols essentially enables independent detections and the sphere decoding complexity is mainly determined by the orthogonality among the first several symbols.
Hence, exploiting the underlying orthogonality in the codeword and carefully choosing the sequence of information symbols can bring benefits in terms of decoding complexity.

\subsection{A new proposal of the 3D MIMO codeword}

The initially proposed codeword matrix of the 3D MIMO code is explicitly written as:
\begin{equation}
\label{eq:3D}
\textbf{X}_{\mathrm{3D}}=\left [\begin{array}{*{20}c}
        \mathbf{X}_{\mathrm{Golden},1} & -\mathbf{X}_{\mathrm{Golden},2}^{\ast}\\
        \mathbf{X}_{\mathrm{Golden},2} & \mathbf{X}_{\mathrm{Golden},1}^{\ast} \\
        \end{array}\right]
=\!\frac{1}{\sqrt{5}}\!\left [\begin{array}{*{20}c}
        \alpha (s_1+\theta s_2) & \alpha (s_3+\theta s_4) & -\alpha^{\ast} (s_5^{\ast}+\theta s_6^{\ast}) & -\alpha^{\ast} (s_7^{\ast}+\theta s_8^{\ast})\\
        i\bar{\alpha} (s_3+\bar{\theta} s_4) & \bar{\alpha} (s_1+\bar{\theta} s_2)  & i\bar{\alpha}^{\ast} (s_7^{\ast}+\bar{\theta} s_8^{\ast}) & -\bar{\alpha}^{\ast} (s_5^{\ast}+\bar{\theta} s_6^{\ast}) \\
        \alpha (s_5+\theta s_6) & \alpha (s_7+\theta s_8) & \alpha^{\ast} (s_1^{\ast}+\theta s_2^{\ast}) & \alpha^{\ast} (s_3^{\ast}+\theta s_4^{\ast})\\
        i\bar{\alpha} (s_7+\bar{\theta} s_8) & \bar{\alpha} (s_5+\bar{\theta} s_6)  & -i\bar{\alpha}^{\ast} (s_3^{\ast}+\bar{\theta} s_4^{\ast}) & \bar{\alpha}^{\ast} (s_1^{\ast}+\bar{\theta} s_2^{\ast}) \\
\end{array}
        \right],
\end{equation}
where $\theta=\frac{1+\sqrt{5}}{2}$, $\bar{\theta}=1-\theta$, $\alpha=1+i(1-\theta)$ and $\bar{\alpha}=1+i(1-\bar{\theta})$.
It is constructed in a hierarchical manner: eight information symbols ($\kappa=8$) are first encoded to two Golden codewords~\cite{belfiore2005golden}, i.e. $\mathbf{X}_{\mathrm{Golden},1}$ and $\mathbf{X}_{\mathrm{Golden},2}$, which are consequently arranged in an Alamouti manner~\cite{alamouti1998simple} over four channel uses ($T=4$)\footnote{Note that this construction is different from that of the quasi-orthogonal code~\cite{sharma2004full} and the EAST code~\cite{sinnokrot2008embedded}.}.
This results in a code rate of $2$ which is \emph{full-rate} for the $4\times2$ MIMO transmission.
Previous study shows that the 3D MIMO code achieves efficient and robust performance.
However, since  eight information symbols are stacked in one codeword, the ML decoding complexity is up to $O(M^8)$.

It was shown that it is possible to achieve lower sphere decoding complexity through permuting the sequence of information symbols~\cite{jithamithra2011minimizing}.
We propose to slightly modify the codeword by exchanging the positions of information symbols $(s_3, s_4)$ and $(s_5,s_6)$, yielding a new form of codeword:
\begin{equation}
\label{eq:3Dnew}
\textbf{X}_{\mathrm{3D},new}= \!\frac{1}{\sqrt{5}}\!\left [\begin{array}{*{20}c}
        \alpha (s_1+\theta s_2) & \alpha (s_5+\theta s_6) & -\alpha^{\ast} (s_3^{\ast}+\theta s_4^{\ast}) & -\alpha^{\ast} (s_7^{\ast}+\theta s_8^{\ast})\\
        i\bar{\alpha} (s_5+\bar{\theta} s_6) & \bar{\alpha} (s_1+\bar{\theta} s_2)  & i\bar{\alpha}^{\ast} (s_7^{\ast}+\bar{\theta} s_8^{\ast}) & -\bar{\alpha}^{\ast} (s_3^{\ast}+\bar{\theta} s_4^{\ast}) \\
        \alpha (s_3+\theta s_4) & \alpha (s_7+\theta s_8) & \alpha^{\ast} (s_1^{\ast}+\theta s_2^{\ast}) & \alpha^{\ast} (s_5^{\ast}+\theta s_6^{\ast})\\
        i\bar{\alpha} (s_7+\bar{\theta} s_8) & \bar{\alpha} (s_3+\bar{\theta} s_4)  & -i\bar{\alpha}^{\ast} (s_5^{\ast}+\bar{\theta} s_6^{\ast}) & \bar{\alpha}^{\ast} (s_1^{\ast}+\bar{\theta} s_2^{\ast}) \\
\end{array}
        \right].
\end{equation}
Since we only change the sequence of the information symbols in the codeword (the third and fourth information symbols become the fifth and sixth, and vice versa) and the information symbols are independent from each other, the new codeword preserves all the good attributes of the original 3D MIMO code in distributed MIMO scenarios illustrated in \cite{liu2013distributed}.
More importantly, this modification is based on the embedded orthogonalities in the 3D MIMO codeword and~\mbox{yields} an interesting codeword structure which will be exploited to achieve lower decoding complexity.
The advantages brought by the new codeword structure will be highlighted in the following sections.
%\sout{as shown in the following sections, }

\subsection{Key properties of the proposed 3D MIMO codeword}
Due to the underlying Alamouti and Golden structures, the 3D MIMO code has some unique properties which lead to simplified decoding algorithms.

For the modified 3D MIMO code (\ref{eq:3Dnew}) over a $4\times2$ MIMO channel, the $\mathbf{R}$ matrix in (\ref{eq:SD_detection_metric}) is a $16\times16$ real-valued matrix. Rewrite $\mathbf{R}$ in a block-wise form:
\begin{equation}\label{eq:R_mat}
 \mathbf{R}=\left [\begin{array}{*{20}c}
  \mathbf{R}_{11} & \mathbf{R}_{12} & \mathbf{R}_{13}  & \mathbf{R}_{14} \\
  0& \mathbf{R}_{22} & \mathbf{R}_{23} & \mathbf{R}_{24} \\
  0 & 0& \mathbf{R}_{33} & \mathbf{R}_{34} \\
  0&0 &0 & \mathbf{R}_{44} \\
 \end{array}\right],
\end{equation}
where $\mathbf{R}_{jk}$'s are $4\times 4$ submatrices containing $\langle \mathbf{q}_m,\mathbf{h}_n\rangle$'s with $m=4(j-1)+1,\ldots,4j$ and $n=4(k-1)+1,\ldots,4k$.

Based on the new codeword in (\ref{eq:3Dnew}) and taking into account (\ref{eq:Heq}), (\ref{eq:ST_coding_by_generator_matrix}) and (\ref{eq:generator_matrix}), we can obtain a few interesting properties of $\mathbf{R}$ that can be made use of to achieve a low decoding complexity.
\begin{theorem}
    \label{theo:R11}
  $\mathbf{R}_{11}$ is an upper triangular matrix with $\langle \mathbf{q}_1,\mathbf{h}_2\rangle=\langle \mathbf{q}_1,\mathbf{h}_4\rangle=\langle \mathbf{q}_2,\mathbf{h}_3\rangle=\langle \mathbf{q}_3,\mathbf{h}_4\rangle=0$. \qed
\end{theorem}
\begin{theorem}
\label{theo:R12}
  $\mathbf{R}_{12}$ is a null matrix when the channel is quasi-static, i.e. $\langle \mathbf{q}_j,\mathbf{h}_k\rangle=0$,  $\forall j=1,2,3,4$ and $k=5,6,7,8$. \qed
\end{theorem}
\begin{corollary}
  \label{cor:R22}
   $\mathbf{R}_{22}$ is an upper triangular matrix with a similar structure as $\mathbf{R}_{11}$, i.e. $\langle \mathbf{q}_5,\mathbf{h}_6\rangle=\langle \mathbf{q}_5,\mathbf{h}_8\rangle=\langle \mathbf{q}_6,\mathbf{h}_7\rangle=\langle \mathbf{q}_7,\mathbf{h}_8\rangle=0$. \qed
\end{corollary}
\begin{figure}[!t]
\centering
\includegraphics[width=2.7in]{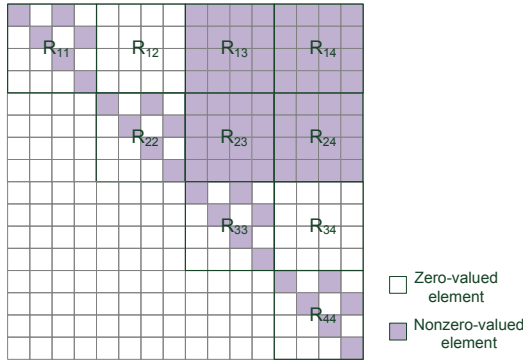}
\caption{Illustration of the $\mathbf{R}$ matrix of the new 3D MIMO codeword (\ref{eq:3Dnew}) in quasi-static channel.}
\label{fig_R_mat_swt}
\end{figure}

The proofs of Theorem~\ref{theo:R11}, Theorem~\ref{theo:R12} and Corollary~\ref{cor:R22} are presented in Appendices.  The above properties are visualized in Figure~\ref{fig_R_mat_swt}.

\textbf{Remark 1}: Theorem~\ref{theo:R11}  and Corollary~\ref{cor:R22} actually suggest the independency between real and imaginary parts of the information symbols.
For instance, $\langle \mathbf{q}_1,\mathbf{h}_2\rangle=\langle \mathbf{q}_1,\mathbf{h}_4\rangle=\langle \mathbf{q}_3,\mathbf{h}_4\rangle=0$ means that the real parts of the first and second received symbols, namely $\widetilde{\mathbf{z}}(1)$ and $\widetilde{\mathbf{z}}(3)$, do not contain any contribution from {$s_1^I$} and {$s_2^I$}.
Similarly, $\langle \mathbf{q}_2,\mathbf{h}_3\rangle=0$ means that their imaginary parts, namely $\widetilde{\mathbf{z}}(2)$ and $\widetilde{\mathbf{z}}(4)$, do not contain any contribution from {$s_1^R$} and {$s_2^R$}, either.
As we will show later, this real/imaginary independency leads to independent and parallel detections for real part and imaginary part, respectively.

The real/imaginary part independency comes from the underlying {Golden and Alamouti structures}.
It has been revealed that the complex-valued $\mathbf{R}$ matrix of the Golden code has a real upper left submatrix~\cite{sinnokrot2010fast}, which coincides with the structure as presented in Theorem~\ref{theo:R11}.
It shows the real/imaginary part independency of the Golden code in its $2\times2$ codeword matrix. The Alamouti-like arrangement of the two Golden codewords, on the other hand, helps creating this independency in the $4\times4$ codeword matrix of the 3D MIMO code.

\textbf{Remark 2}: Theorem~\ref{theo:R12} indicates that the information symbols  $s_1$ and $s_2$ are uncorrelated with  $s_3$ and $s_4$ in the received symbols.
It means that these two symbol groups can be determined independently. We
only have to jointly determine six information symbols to obtain the ML solutions.
In other words, the ML decoding complexity is expected to be $O(M^6)$ instead of $O(M^8)$.
Therefore, \emph{the 3D MIMO code is fast decodable}.
Details on the simplified ML decoding will be discussed in Section~\ref{sec:MLdec}.

%the first two received complex symbols, or equivalently $\widetilde{\mathbf{z}}(1)$, $\widetilde{\mathbf{z}}(2)$, $\widetilde{\mathbf{z}}(3)$ and $\widetilde{\mathbf{z}}(4)$, do not contain any contribution from information symbols $s_3$ and $s_4$.
%Hence, a group of six information symbols $s_1$, $s_2$, $s_5$, $s_6$, $s_7$, $s_8$ can be jointly determined regardless the values of $s_3$ and $s_4$.
%It means that the ML decoding can be achieved by joint searches over six, instead of eight, information symbols.

It should be noted that Theorem~\ref{theo:R12} is partially enabled by the embedded Alamouti structure in the codeword.
The channel coefficients should be constant within the duration of one codeword to validate the orthogonalities in Alamouti structure.
Hence Theorem~\ref{theo:R12} is only valid in the quasi-static channels\footnote{The fast decodability of other STBCs such as  BHV, Srinath-Rajan, IFS, also requires quasi-static channel assumption.}.

\subsection{Comparison with the original 3D MIMO codeword}
\begin{figure}[!t]
\centering
\includegraphics[width=2.7in]{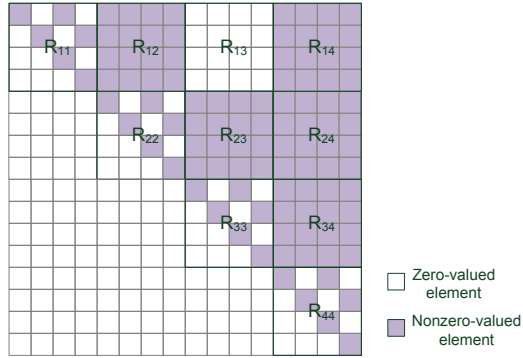}
\caption{Illustration of the $\textbf{R}$ matrix of the original 3D MIMO codeword given in (\ref{eq:3D}) in quasi-static channel.}
\label{fig_R_mat}
\end{figure}

Figure~\ref{fig_R_mat_swt} illustrates the $\mathbf{R}$ matrix of the new 3D MIMO codeword.
Compared with the original one as shown in Figure~\ref{fig_R_mat}, the new structure is actually more favorable for the MIMO decoding.
In the new codeword, the contributions of information symbol groups  $(s_1, s_2)$ and $(s_3, s_4)$ are totally uncorrelated in the received signal, which means that the ML detection of eight information symbols can be achieved by two independent and less complex detections of six information symbols.
Moreover, the structure of $\mathbf{R}_{11}$ and $\mathbf{R}_{22}$ enables the independent detections of real and imaginary parts of $(s_1, s_2)$ and $(s_3, s_4)$, which leads to further complexity reduction.
Yet, this real/imaginary parts separation is not straightforward in the original codeword.

It should be emphasized that the new codeword only changes the sequence of the information symbols in the codeword to facilitate the decoding process.
It does not affect all the good properties of the 3D MIMO code.

\section{Proposed ML Decoder with Low Complexity}
\label{sec:MLdec}

In this section, a low-complexity ML decoding algorithm exploiting the unique properties highlighted in previous section is proposed for the 3D MIMO code.
Generally speaking, the complexity reduction is achieved in two steps.
Based on Theorem~\ref{theo:R12} the joint detection of eight information symbols is converted into two partially independent detections of six information symbols.
This step reduces the worst-case decoding complexity from $O(M^8)$ to $O(M^{6})$.
Consequently, using Theorem~\ref{theo:R11} and Corollary~\ref{cor:R22}, the detections of complex information symbols are converted into independent detections of real and imaginary parts, which further reduces the worst-case complexity to $O(M^{4.5})$.
%The simplified ML decoder will be formulated in the following part.

\subsection{Group-wise parallel detections}

We divide the information symbols and received symbols into four groups, i.e. $\mathbf{a}=\widetilde{[s_1, s_2]^T}$, $\mathbf{b}=\widetilde{[s_3, s_4]^T}$, $\mathbf{c}=\widetilde{[s_5, s_6]^T}$, $\mathbf{d}=\widetilde{[s_7, s_8]^T}$, $\mathbf{z}_{12}=\widetilde{[z_1, z_2]^T}$, $\mathbf{z}_{34}=\widetilde{[z_3, z_4]^T}$, $\mathbf{z}_{56}=\widetilde{[z_5, z_6]^T}$ and $\mathbf{z}_{78}=\widetilde{[z_7, z_8]^T}$.
Taking into account the structure of $\mathbf{R}$ and Theorem~\ref{theo:R12}, the decoding metric in (\ref{eq:SD_detection_metric}) can be rewritten as:
\begin{align}
\|\widetilde{\mathbf{z}}-\mathbf{R}\widetilde{\mathbf{s}}\|^2& = \|\mathbf{z}_{12}-\mathbf{R}_{11}\mathbf{a}-\mathbf{R}_{13}\mathbf{c}-\mathbf{R}_{14}\mathbf{d} \|^2 \label{eq:ML_metric_original1}\\
&+ \|\mathbf{z}_{34}-\mathbf{R}_{22}\mathbf{b}-\mathbf{R}_{23}\mathbf{c}-\mathbf{R}_{24}\mathbf{d} \|^2 \label{eq:ML_metric_original2}  \\
&+ \|\mathbf{z}_{56}-\mathbf{R}_{33}\mathbf{c}-\mathbf{R}_{34}\mathbf{d} \|^2  + \|\mathbf{z}_{78}-\mathbf{R}_{44}\mathbf{d} \|^2. \nonumber
\end{align}
From (\ref{eq:ML_metric_original1}) and (\ref{eq:ML_metric_original2}), it can be seen that the contributions from the information symbol groups $\mathbf{a}$ and $\mathbf{b}$ are uncorrelated in the received symbol.
For instance, $\mathbf{z}_{12}$ does not contain any information from $\mathbf{b}$, and $\mathbf{z}_{34}$ is irrelevant to $\mathbf{a}$, either.
This enables us to use group-wise conditional detections to retrieve the ML solutions~\cite{sirianunpiboon2010fast}.

In particular, {the ML solution $\hat{\mathbf{s}}^{\mathrm{ML}}=[\hat{\mathbf{a}}, \hat{\mathbf{b}}, \hat{\mathbf{c}}, \hat{\mathbf{d}}]^T$ is achieved in two search steps, namely a joint ``outer''} {search for $[\hat{\mathbf{c}}, \hat{\mathbf{d}}]$}:
\begin{align}
\label{eq:min_cd}
  [\hat{\mathbf{c}}, \hat{\mathbf{d}}]\!&=\!\arg\!\!\min_{[\mathbf{c},\mathbf{d}]\in\boldsymbol\Theta^4}\!\! \big( \|\mathbf{v}_{12}-\mathbf{R}_{11}{\mathbf{a}}^{(\ast)}(\mathbf{c}, \mathbf{d}) \|^2 + \|\mathbf{v}_{34} - \mathbf{R}_{22}{\mathbf{b}}^{(\ast)}(\mathbf{c}, \mathbf{d}) \|^2 +\|\mathbf{z}_{56}-\mathbf{R}_{33}\mathbf{c}-\mathbf{R}_{34}\mathbf{d} \|^2+\|\mathbf{z}_{78}-\mathbf{R}_{44}\mathbf{d} \|^2 \big),
\end{align}
{and two independent ``inner'' searches for $\hat{\mathbf{a}}$ and $\hat{\mathbf{b}}$, respectively:}
\begin{align}
  \hat{\mathbf{a}} = {\mathbf{a}}^{(\ast)}(\hat{\mathbf{c}}, \hat{\mathbf{d}}), \quad \hat{\mathbf{b}} = {\mathbf{b}}^{(\ast)}(\hat{\mathbf{c}}, \hat{\mathbf{d}})
\end{align}
where
\begin{align}
 {\mathbf{a}}^{(\ast)}(\mathbf{c}, \mathbf{d}) &= \arg\min_{\mathbf{a}\in\boldsymbol\Theta^2}\|\mathbf{v}_{12}-\mathbf{R}_{11}\mathbf{a} \|^2, \label{eq:min_a} \\
 {\mathbf{b}}^{(\ast)}(\mathbf{c}, \mathbf{d}) &= \arg \min_{\mathbf{b}\in\boldsymbol\Theta^2} \|\mathbf{v}_{34}- \mathbf{R}_{22}\mathbf{b} \|^2,\label{eq:min_b}
\end{align}
with $\mathbf{v}_{12}=\mathbf{z}_{12}-\mathbf{R}_{13}\mathbf{c}-\mathbf{R}_{14}\mathbf{d}$,
$\mathbf{v}_{34}=\mathbf{z}_{34}-\mathbf{R}_{23}\mathbf{c}-\mathbf{R}_{24}\mathbf{d}$.
The outer search is carried out over the combinations of $[\mathbf{c},\mathbf{d}]$.
For a given $[\mathbf{c},\mathbf{d}]$, the search of $\mathbf{a}$ and the search of $\mathbf{b}$ are performed in parallel.
The concatenation of outer and inner searches (either $\mathbf{a}$ or $\mathbf{b}$) results in a joint search of six information symbols.
Therefore, the worst-case decoding complexity is reduced from $O(M^8)$ to $O(M^6)$.
We note that this complexity reduction does not rely on the constellation that is adopted by the information symbols.
In other words, the 3D MIMO code requires a worst decoding complexity of $O(M^6)$ for arbitrary modulation.

\subsection{Independent detections of real and imaginary parts}
If square shape QAM modulations are considered, the decoding complexity can be further improved.
The square $M$-QAM symbol can be separated into two independent $\sqrt{M}$-PAM symbols on the real and imaginary axes, respectively.
Using Theorem~\ref{theo:R11} and Corollary~\ref{cor:R22}, the real and imaginary parts can be decoded separately.
Take the detection of $\mathbf{a}$ as an example.
Denote its real and imaginary parts as $\mathbf{a}^R=[s_1^R,s_2^R]^T$ and $\mathbf{a}^I=[s_1^I,s_2^I]^T$, respectively.
Given $[\mathbf{c},\mathbf{d}]$ and using Theorem~\ref{theo:R11}, the detection of $\mathbf{a}$ in (\ref{eq:min_a}) is rewritten as~\cite{sinnokrot2010fast}:
%\begin{align}\label{eq:dec_a}
%\arg\min_{\mathbf{a}\in\boldsymbol\Theta^2}\|\mathbf{v}_{12}-\mathbf{R}_{11}\mathbf{a} \|^2=
%     \arg\!\!\min_{\mathbf{a}^R\in\boldsymbol\Psi^2}\!\!\|\mathbf{v}_{12}^R-\mathbf{R}_{11}^R\mathbf{a}^R \|^2 \! + \!    \arg\!\!\min_{\mathbf{a}^I\in\boldsymbol\Psi^2}\!\!\|\mathbf{v}_{12}^I-\mathbf{R}_{11}^I\mathbf{a}^I \|^2,
%\end{align}
\begin{align}\label{eq:dec_a}
     \hat{\mathbf{a}}^R = \arg\min_{\mathbf{a}^R\in\boldsymbol\Psi^2}\|\mathbf{v}_{12}^R-\mathbf{R}_{11}^R\mathbf{a}^R \|^2, \quad \hat{\mathbf{a}}^I = \arg\min_{\mathbf{a}^I\in\boldsymbol\Psi^2}\|\mathbf{v}_{12}^I-\mathbf{R}_{11}^I\mathbf{a}^I \|^2,
\end{align}
where $\boldsymbol\Psi$ is the set of $\sqrt{M}$-PAM constellation symbols; $\mathbf{v}^R_{12}=[v_1^R,v_2^R]^T$, $\mathbf{v}^I_{12}=[v_1^I,v_2^I]^T$; $\mathbf{R}_{11}^R$ and $\mathbf{R}_{11}^I$ are tailored upper-triangular matrices associated with real and imaginary parts, respectively:
\begin{align}
  \mathbf{R}_{11}^R=\left [\begin{array}{*{20}c}
  \mathbf{R}_{11}(1,1)& \mathbf{R}_{11}(1,3)\\
  0& \mathbf{R}_{11}(3,3) \\
  \end{array}\right], \quad \mathbf{R}_{11}^I=\left [\begin{array}{*{20}c}
  \mathbf{R}_{11}(2,2)& \mathbf{R}_{11}(2,4)\\
  0& \mathbf{R}_{11}(4,4)\\
  \end{array}\right].
\end{align}
(\ref{eq:dec_a}) means that the detections of real and imaginary parts are similar and can be performed separately.
Take the real part as an example.
We apply again the conditional detection here.
For a given $s_2^R$, the metric for the real part detection becomes:
\begin{equation}\label{eq:metric_a_R}
    \|\mathbf{v}_{12}^R-\mathbf{R}_{11}^R\mathbf{a}^R \|^2\! = \! \big(w_{1}^R-\mathbf{R}_{11}(1,1)s_1^R\big)^2 + (w_2^R)^2,
\end{equation}
where $w_{1}^R=v_1^R-\mathbf{R}_{11}(1,3)s_2 ^R$ and $w_2^R=v_2^R-\mathbf{R}_{11}(3,3)s_2 ^R$.
For a given $s_2^R$, the best $s_1^R$ that minimizes the decoding metric can alternatively be found by minimizing a quadratic function of $s_1^R$ given on the right hand side of (\ref{eq:metric_a_R}).
The best solution of $s_1^R$ is easily found by:
\begin{equation}\label{eq:s_1R}
    \overline{s}_1^R=\texttt{Q}\Big(\frac{v_1^R-\mathbf{R}_{11}(1,3)s_2^R}{\mathbf{R}_{11}(1,1)}\Big),
\end{equation}
where $\texttt{Q}(\cdot)$ is the slicing operation providing the PAM symbol that is closest to the given value.
The best combination of $[\hat{s}_1^R,\hat{s}_2^R]^T$ given $[\mathbf{c},\mathbf{d}]$ is obtained after testing (\ref{eq:s_1R}) with all ($\sqrt{M}$) possible values of $s_2^R$:
\begin{align}
    \label{eq:s_1_s_2R}
  \hat{s}_2^R= \arg\min_{s_2^R\in\boldsymbol\Psi}\big(|v_1^R-\mathbf{R}_{11}(1,1)\overline{s}_1^R-\mathbf{R}_{11}(1,3)s_2^R|^2
+|v_2^R-\mathbf{R}_{11}(3,3)s_2^R|^2\big).
\end{align}
{Consequently, $\hat{s}_1^R$ is obtained by using the solution $\hat{s}_2^R$ in (\ref{eq:s_1R}).
}
Similar process can be applied to solve the imaginary parts.
The best solution of $[\hat{s}_1^I,\hat{s}_2^I]^T$ given $[\mathbf{c},\mathbf{d}]$ can be found by:
\begin{align}
 \label{eq:s_1_s_2I}
  \hat{s}_2^I = \arg\min_{{s}_2^I\in\boldsymbol\Psi}\big(|v_1^I-\mathbf{R}_{11}(2,2)\overline{s}_1^I-\mathbf{R}_{11}(2,4){s}_2^I|^2+|v_2^I-\mathbf{R}_{11}(4,4){s}_2^I|^2\big),
\end{align}
where
\begin{equation}\label{eq:s_1I}
    \overline{s}_1^I=\texttt{Q}\Big(\frac{v_1^I-\mathbf{R}_{11}(2,4){s}_2^I}{\mathbf{R}_{11}(2,2)}\Big).
\end{equation}
{$\hat{s}_1^I$ is computed by applying the solution $\hat{s}_2^R$ in (\ref{eq:s_1I}).}

Using the same technique, the best solutions of $\mathbf{b}$ in (\ref{eq:min_b}) can also be converted into independent detections of $\mathbf{b}^R$ and $\mathbf{b}^I$.
Substituting $\mathbf{R}_{11}$, $v_1$, $v_2$, $s_1$ and $s_2$ in (\ref{eq:s_1R}), (\ref{eq:s_1_s_2R}), (\ref{eq:s_1_s_2I}) and (\ref{eq:s_1I}) by $\mathbf{R}_{22}$, $v_3$, $v_4$, $s_3$ and $s_4$, respectively, it yields the detections for $s_3$ and $s_4$.
In general, for a given $[\mathbf{c},\mathbf{d}]$, the search of two complex symbols $[\mathbf{a}, \mathbf{b}]$ is turned into four independent searches of $\sqrt{M}$ PAM symbols.
The resulting overall complexity to decode a whole codeword is $O(M^{4.5})$.

\begin{table}[!t]
\renewcommand{\arraystretch}{1.2}
\caption{Comparison of ML decoding complexities of full-rate STBCs for $4\times2$ MIMO transmission}
\label{tbl:complexity}
\centering
        \renewcommand{\arraystretch}{1.2}
        \begin{tabular}{|c|c|c|}
            \hline
            \multirow{2}{*}{\textbf{STBC}} & \multicolumn{2}{c|}{\textbf{ML decoding complexity}}\\ \cline{2-3}
                & any QAM & square QAM \\ \hline
            new 3D MIMO & $O(M^6)$ & $O(M^{4.5})$\\ \hline
            DjABBA~\cite{hotinen2003multiantenna} & $O(M^7)$ & $O(M^{6})$\\ \hline
            Perfect code (2-layer)~\cite{oggier2006perfect} & $O(M^6)$ & $O(M^{5.5})$ \\ \hline
            BHV~\cite{biglieri2009fast} & $O(M^6)$ & $O(M^{4.5})$\\ \hline
            EAST~\cite{sinnokrot2008embedded} & $O(M^5)$ & $O(M^{4.5})$\\ \hline
            Srinath-Rajan~\cite{srinath2009low} & $O(M^5)$ & $O(M^{4.5})$\\ \hline
            IFS~\cite{ismail2013new} & $O(M^5)$ & $O(M^{4.5})$\\ \hline
        \end{tabular}
\end{table}

In summary, the 3D MIMO code requires a worst decoding complexity of $O(M^6)$ for any modulation scheme and $O(M^{4.5})$ for square $M$-QAM modulations.
Recall that the 3D MIMO code achieves a coding rate of 2 which is full-rate for $4\times2$ MIMO transmissions.
The comparisons with other state-of-the-art full-rate STBCs are presented in Table 1.
It can be seen that the 3D MIMO code is among the simplest full-rate STBCs when the square QAM modulations are considered.

\section{Proposed Implementation of the Simplified ML Decoder}
\label{sec:ImpMLdec}
In the previous sections, we have illustrated the fast decodability of the 3D MIMO code in theory.
With this knowledge, we propose an implementation of the simplified ML decoder that can be used in practice.
Using the two-stage tree-search structure and leveraging the symmetry structure in the codeword, the proposed implementation requires a low average complexity in practice.
Moreover, various performance-complexity trade-offs can be easily achieved by replacing the sphere decoder by other suboptimal tree search algorithms such as K-best algorithm~\cite{wong2002vlsi}, fixed-complexity sphere decoder~\cite{barbero2008fixing} etc.

\subsection{Two-stage decoding structure}

Recall that the fast decodability is achieved by concatenating the joint search of four complex symbols and several detections in parallel.
Figure~\ref{fig_2step_dec} presents the general structure of the proposed simplified ML decoder.
Detailed pseudo code is presented in Algorithms~\ref{algo_main}, \ref{algo_simp_tree}, \ref{algo_para_dec} and \ref{algo_col_swt} so that the proposed decoder can be implemented without major effort.

\begin{figure}[!t] % figure 3
\centering
\includegraphics[width=3.4in]{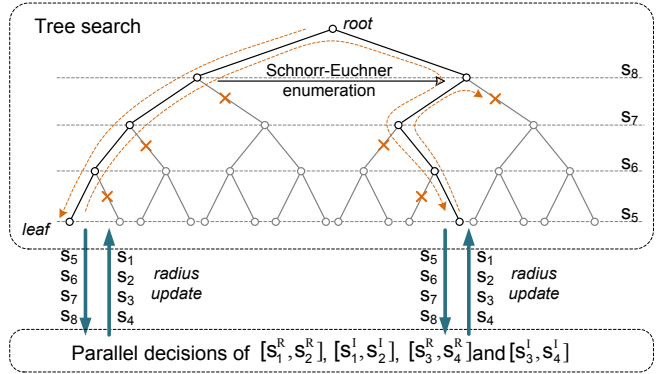}
\caption{Illustration of the  two-stage sphere decoding.}
\label{fig_2step_dec}
\end{figure}

\subsubsection{4-level tree search phase}
The joint detection of $[\mathbf{c}, \mathbf{d}]$ is realized by a complex sphere decoder with Schnorr-Euchner enumeration, which is visualized by the search over a 4-level tree as shown in Figure~\ref{fig_2step_dec}.
The nodes of the same level represent all the solutions of a complex information symbol.
Each path from the root to a leaf node represents a possible combination of $[\mathbf{c}, \mathbf{d}]$.

The details of the tree search is explicitly presented in Algorithm~\ref{algo_simp_tree}.
The search starts from the root node and traverses the nodes of lower levels in a depth-first manner.
An adaptive search radius is used to speed up the convergence of the algorithm by limiting the search within a hypersphere $\mathcal{S}$.
For the node under checking, the partial distance resulted by the current path is compared with the radius.
If the partial distance is smaller than the radius, the search moves on to the children nodes on the next level. Otherwise, the search jumps to another sibling node on the current level.
When all the nodes of the level have already been checked, the search goes back to the upper level.
The radius is initially set to infinity and is adaptively decreased according to the best solution already found in the search.
Specifically, the radius is updated taking into account the best combination of $[\mathbf{c}, \mathbf{d}]$ and $[\mathbf{a}, \mathbf{b}]$ (line~\ref{algo_radius_new} of Algorithm~\ref{algo_simp_tree}).
The latter is obtained from the parallel decisions phase.
The tree search is terminated when all nodes within the hypersphere have been checked.
The best solution is the ML solution.

The sequence in which the sibling nodes are visited is determined according to the their partial distances in an ascending order.
This is to guarantee that the promising candidates are visited first in order to reduce the search complexity.
This ordering process is referred to as Schnorr-Euchner enumeration~\cite{schnorr1994lattice,agrell2002closest,guo2004reduced}.
It can simply be implemented by a lookup table~\cite{wiesel2003efficient,tsai2010a4x4} (line~\ref{algo_SE} in Algorithm~\ref{algo_main}) and its complexity is merely the computation of the linear estimation $\hat{s}_{_{\mathrm{ZF}}}$.

\subsubsection{Parallel decision phase}

Once a leaf node is achieved in the tree search, a better solution of  $[\mathbf{c}, \mathbf{d}]$  is found.
Consequently, the tree search process is suspended and  the new $[\mathbf{c}, \mathbf{d}]$ is used to trigger the parallel detections of rest symbols.

\begin{figure}[!t] % figure 4
\centering
\includegraphics[width=3.4in]{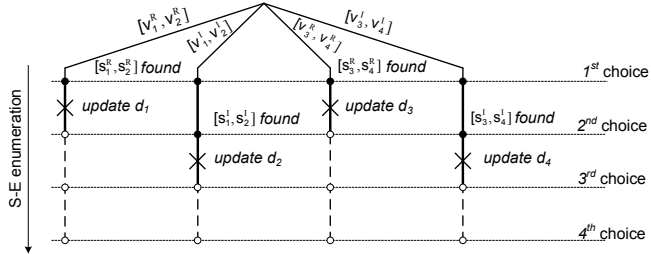}
\caption{Parallel decisions of $[s_1^R,s_2^R]$, $[s_1^I,s_2^I]$, $[s_3^R,s_4^R]$ and $[s_3^I,s_4^I]$.}
\label{fig_parallel_dec}
\end{figure}

The parallel detection is depicted in Figure~\ref{fig_parallel_dec}. The implementation details are presented in Algorithm~\ref{algo_para_dec}.
As shown in Figure~\ref{fig_parallel_dec}, the detections of $[s_1^R,s_2^R]$, $[s_1^I,s_2^I]$, $[s_3^R,s_4^R]$ and $[s_3^I,s_4^I]$ are carried out in parallel.
For each branch, a one-level sphere decoder is used to traverse all possible PAM symbols as given in (\ref{eq:s_1_s_2R}).
The visiting sequence is also determined by the Schnorr-Euchner enumeration.
The detections in different branches are synchronized by a common clock signal because the operations are exactly the same for all branches.
All branches simultaneously check the first candidate PAM symbol and then move on to the second one, and so on.

Moreover, we propose a mechanism that terminates the search in each branch not only based on its own results, but also taking into account the results from other branches.
In particular, once the best solution of the $j$th branch is found ahead of others, the resulting branch distance $d_j$ is recorded and shared with other branches to speed up the overall search process.

Take the search of the first branch as an example.
The most promising PAM symbol in the unchecked symbol list is assigned to $\overline{s}_2^R$ (line~\ref{algo_s2R} in Algorithm~\ref{algo_para_dec}).
The partial distance $\tau_1$ is calculated (line~\ref{algo_tau1} in Algorithm~\ref{algo_para_dec}).
The search is terminated in two cases: 1) if this partial distance is greater than the current minimum branch distance ($\tau_1>p_1$); or 2) if the overall distance is beyond the current radius of the sphere decoder in the tree search phase ($(\tau_1+d_2+d_3+d_4+d)>\mathrm{radius}$).

Once the searches on all branches are terminated, the solution $[\mathbf{a}, \mathbf{b}]$ and the resulting distance $d_p$ are returned to the tree search phase.
The tree search process is resumed.
The overall distance is compared with the current radius (line~\ref{algo_radius_cmp} in Algorithm~\ref{algo_simp_tree}) to determine whether the current solution is a better one.
If a better solution is found, the radius is updated accordingly (line~\ref{algo_radius_new} in Algorithm~\ref{algo_simp_tree}).
The tree search process is moved on to the next unchecked node.

\subsection{Column switch based on ZF estimation}

In the proposed algorithm, the search of eight symbols is divided into a tree search for four symbols and parallel detections for the other four symbols.
Due to the symmetric structure of  the codeword matrix (\ref{eq:3Dnew}), some parts of the codewords can be exchanged without changing the properties of the 3D MIMO code.
For instance, we have the same properties as illustrated in Section~\ref{sec:3DMIMO} after exchanging the positions of $[s_1,s_2,s_3,s_4]$ with $[s_5,s_6,s_7,s_8]$.
Similarly, if we exchange $[s_1,s_2]$ with $[s_3,s_4]$ and exchange $[s_5,s_6]$ with $[s_7,s_8]$ simultaneously, the structure of $\mathbf{R}$ matrix maintains, as well.
That is to say, besides the original symbol sequence, the proposed low-complexity decoding algorithm is also valid with other three permuted symbol sequences, i.e. $[s_5,s_6,s_7,s_8,s_1,s_2,s_3,s_4]$, $[s_3,s_4,s_1,s_2,s_7,s_8,s_5,s_6]$ and       $[s_7,s_8,s_5,s_6,s_3,s_4,s_1,s_2]$.

The exchanging of the symbol sequences can be achieved by permuting the corresponding columns in the equivalent channel matrix $\mathbf{H}_{eq}$.
Note that, the aforementioned column permutations do not affect the decoding performance.
This permits us to choose the symbols that will be determined by the tree search and the ones that will be decoded in the parallel detections.

The proposed column switch method is presented in Algorithm~\ref{algo_col_swt}.
The basic idea is to use the tree search to determine the more difficult half part and use the parallel detections to find the easier half part.
The reason behind this idea is that the parallel decoding is more efficient to decode the reliable symbols separately.
The more accurate the linear estimation, the faster convergence speed for each individual detection branch.
On the other hand, the tree search phase is a joint serial detection in nature which is more suitable to decode those unreliable symbols.

The next question is how to properly choose the unreliable symbols.
In the literature, Barbero \emph{et al.} proposed to sort the decoding sequence based on the norm of subchannels in the fixed-complexity sphere decoder\cite{barbero2008fixing}. However, it is not applicable here because the 3D MIMO code achieves full-diversity and the equivalent subchannels have similar norm values.
In addition, as we have to maintain the structure of $\mathbf{R}$ matrix, the unconstrained subchannel sorting proposed in~\cite{wiesel2003efficient} is not applicable, either.

\begin{figure}[!t] % figure5
\centering
\includegraphics[width=3in]{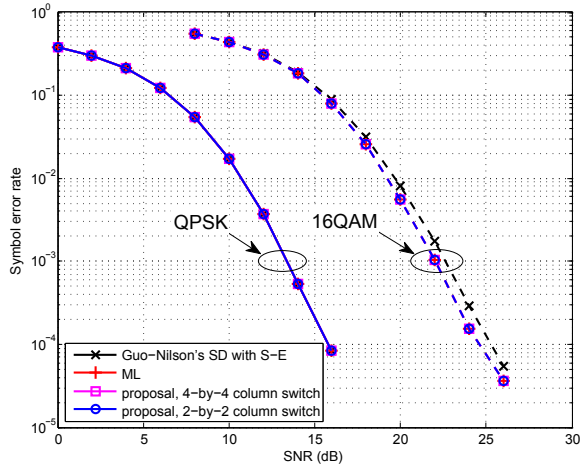}
\caption{Uncoded symbol error rate obtained with Guo-Nilson's sphere decoder with S-E for 3D MIMO code, ML decoder and proposed simplified ML decoders in quasi-static Rayleigh channel with QPSK and 16QAM.}
\label{fig_SER_comparison}
\end{figure}

Alternatively, we propose to sort the information symbols according to the aggregate error of the linear estimation:
\begin{equation}
\label{eq:sort_metric}
  \epsilon_{jk} = \sum_{l=j}^{k}| \hat{\mathbf{s}}_{_{\mathrm{ZF}}} (l)- \mathbf{s}_{_{\mathrm{ZF}}}(l)|^2,
\end{equation}
where $\mathbf{s}_{_{\mathrm{ZF}}}=\mathbf{H}_{eq}^{\dagger}\mathbf{y}$ is the unconstrained estimation of the information symbols in which $\mathbf{H}_{eq}^{\dagger}$ represents the inverse of equivalent channel matrix; $\hat{\mathbf{s}}_{_{\mathrm{ZF}}}=\texttt{Q}(\mathbf{s}_{_{\mathrm{ZF}}})$ is the constellation point that is closest to $\mathbf{s}_{_{\mathrm{ZF}}}$.
The metric is the distance between the estimated information symbols and the nearest constellation points, i.e. an indicator of the estimation accuracy.

Using (\ref{eq:sort_metric}), the decoding sequence can be determined in two levels.
We first compare the aggregate errors of the first half and second half parts of symbols (line~\ref{algo_cmp_2half} in Algorithm~\ref{algo_col_swt}). The half with worse accuracy is assigned to the tree search  (put in the latter part of the decoding sequence).
Consequently, within this half part, the errors of the first two symbols and the second two are compared.
The two symbols with worse accuracy are put closer to the root of the tree.
If this two-symbol by two-symbol exchange takes place in the second half of the symbols which are to be decoded using tree search, the same two-symbol by two-symbol exchange should be done accordingly in the other half in order to maintain the structure of $\mathbf{R}$ matrix.
If only the symbol exchange between two halves of the symbols is carried out, it is referred to as ``4-by-4 column switch''. Otherwise, if the exchange within each half is also performed, it is called ``2-by-2 column switch''.
The advantage of the column switch will be shown in the next section.

\section{Simulation Results}
\label{sec:simu}
%\subsection{Simulation settings}
%The performance is evaluated in the quasi-static independent Rayleigh flat fading channel.

\subsection{BER performance}

Figure~\ref{fig_SER_comparison} presents the uncoded symbol error rate of the proposed simplified decoders in quasi-static independent Rayleigh flat fading channel.
The performances of the ML decoder and the sphere decoder with Schnorr-Euchner (S-E) enumeration proposed by Guo and Nilson~\cite{guo2004reduced} are also given as references.
The Guo-Nilson's sphere decoder is a low-complexity implementation of sphere decoder with S-E but is sub-optimal in terms of symbol error rate.
It can be seen that the proposed decoders achieve the same performance as ML decoder with both QPSK and 16-QAM modulations.
In addition, the proposed decoders outperform the Guo-Nilson's sphere decoder with 16-QAM modulation.
A gain of around $0.7$~dB can be observed at symbol error rate level of $1\times 10^{-4}$.

\subsection{Computational complexity}

\begin{figure}[!t] % figure 6
\centering
\includegraphics[width=3in]{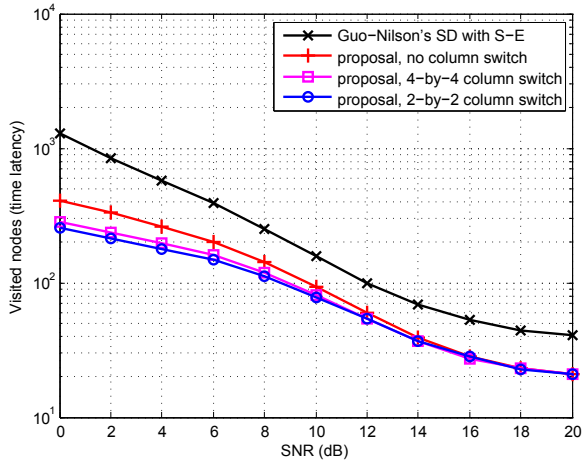}
\caption{Computational complexity in terms of visited nodes required by
 Guo-Nilson's sphere decoder with S-E for 3D MIMO code and proposed simplified ML decoders, in quasi-static Rayleigh channel with QPSK modulation.}
\label{fig_visited_nodes_QPSK}
\end{figure}

\begin{figure}[!t] % figure 7
\centering
\includegraphics[width=3in]{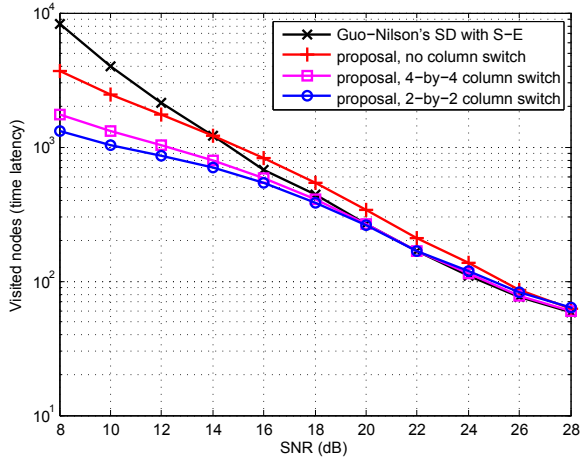}
\caption{Computational complexity in terms of visited nodes required by
 Guo-Nilson's sphere decoder with S-E for 3D MIMO code and proposed simplified ML decoders, in quasi-static Rayleigh channel with 16-QAM modulation.}
\label{fig_visited_nodes_16QAM}
\end{figure}

Figure~\ref{fig_visited_nodes_QPSK} and Figure~\ref{fig_visited_nodes_16QAM} present the complexity in terms of number of visited nodes for decoding each codeword with QPSK and 16-QAM, respectively.
For the proposed decoders, this number is calculated as the number of visited nodes in the tree search phase plus the maximum visited nodes among the four search branches.
The Guo-Nilson's sphere decoder is also given as a reference.
Since processing each node requires roughly the same operations for both decoders, these experiments actually give the comparison of the processing time latency~\cite{guo2004reduced,sinnokrot2010fast}.
It can be seen from the results that the proposed decoders require much less processing time than the ML decoder which needs to traverse all $M^8$ possibilities.
In QPSK case, the proposed decoders always yield less latency than Guo-Nilson's sphere decoder.
For instance, the proposed decoder with 2-by-2 column switch visits only $254.6$ nodes on an average at SNR of $0$~dB.
Compared with the Guo-Nilson's decoder which visits $1276.6$ nodes at SNR of $0$~dB, the proposed one achieves a processing time reduction of $80\%$.
The reductions are $50\%$ and $49\%$ at $10$~dB and $20$~dB, respectively.
In addition, the improvements brought by the proposed column switch technique can also be seen in the results.
For instance, the 2-by-2 column switch yields a processing time reduction of $38\%$ at $0$~dB compared with the decoder without column switch.
This improvement is less significant in high SNR region (e.g. greater than $15$~dB).
Moreover, the 2-by-2 column switch offers better performance than the 4-by-4 counterpart because it is more likely to allocate the four most unreliable complex symbols to the tree-search stage, which helps improving the global convergence speed.
In 16-QAM case (see Figure~\ref{fig_visited_nodes_16QAM}), the proposed decoder with 2-by-2 column switch also brings processing time reduction in low SNR region.
At $8$~dB, it visits $1301.1$ nodes on an average, yielding a time reduction of $84\%$ compared with Guo-Nilson's decoder.
The proposed decoder needs similar time latency as Guo-Nilson's decoder in higher SNR region (e.g. greater than $15$~dB).
Taking into account the symbol error rate performance given in Figure~\ref{fig_SER_comparison}, $8\sim15$ dB is the SNR region where the error correction ability of the channel coding will be carried out significantly.
% That is to say, the improvements are achieved in a SNR region of interest.
This region also represents the minimum SNR level at which the receiver can still work properly.

\begin{figure}[!t] % figure 8
\centering
\includegraphics[width=3in]{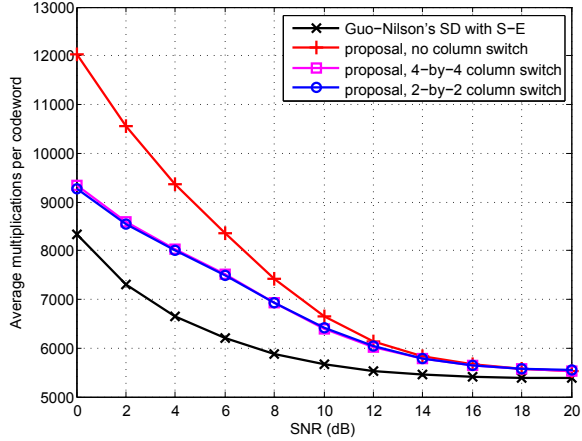}
\caption{Computational complexity in terms of multiplications required by
 Guo-Nilson's sphere decoder with S-E for 3D MIMO code and proposed simplified ML decoders, in quasi-static Rayleigh channel with QPSK modulation.}
\label{fig_multiplication_QPSK}
\end{figure}

\begin{figure}[!t] % figure 9
\centering
\includegraphics[width=3in]{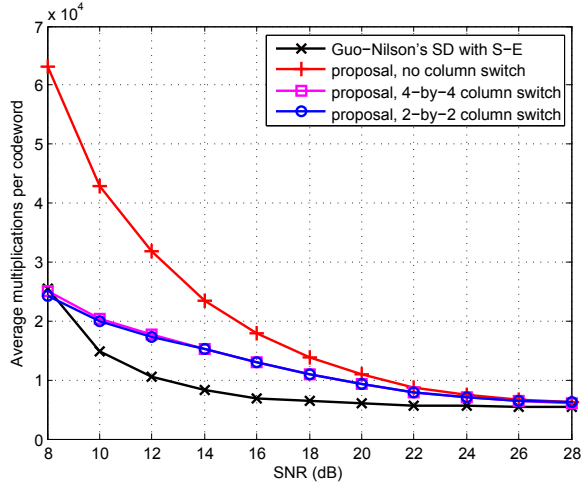}
\caption{Computational complexity in terms of multiplications required by
 Guo-Nilson's sphere decoder with S-E for 3D MIMO code and proposed simplified ML decoders, in quasi-static Rayleigh channel with 16-QAM modulation.}
\label{fig_multiplication_16QAM}
\end{figure}

Figure~\ref{fig_multiplication_QPSK} and Figure~\ref{fig_multiplication_16QAM} give the overall required multiplications to decode each codeword.
For the proposed decoders, the multiplications spent by the tree search and by all four search branches are taken into account.
The computation overheads such as QR decomposition, linear estimation, are also included in the results to give the overall complexity of the decoders.
It can be seen from the results that, in QPSK case, the proposed decoder with 2-by-2 column switch spends $11\%$, $21\%$ and $3\%$ more multiplications than Guo-Nilson's decoder at SNR of $0$~dB, $6$~dB and $20$~dB, respectively.
In 16-QAM case, it needs $5\%$ less multiplication at $8$~dB but spends $85\%$ and $9\%$ more multiplications at $14$~dB and $28$~dB, respectively.
However, it is worth noting that with the cost of increased multiplications the proposed decoders provide less processing latencies.
For instance, the proposed decoder with 2-by-2 column switch achieves $62\%$ processing time reduction at SNR of $6$~dB with QPSK and $42\%$ reduction at $14$~dB with 16-QAM, respectively.

\begin{figure}[!t] % figure 10
\centering
\includegraphics[width=3in]{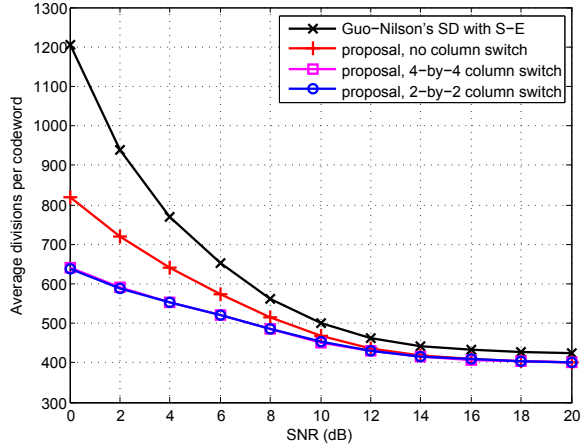}
\caption{Computational complexity in terms of divisions required by
 Guo-Nilson's sphere decoder with S-E for 3D MIMO code and proposed simplified ML decoders, in quasi-static Rayleigh channel with QPSK modulation.}
\label{fig_division_QPSK}
\end{figure}

\begin{figure}[!t]  % figure 11
\centering
\includegraphics[width=3in]{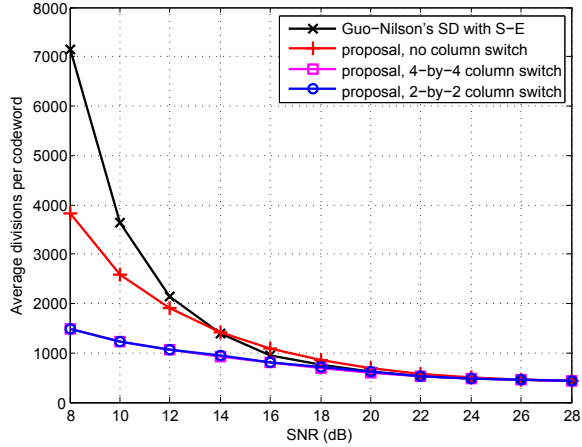}
\caption{Computational complexity in terms of divisions required by
 Guo-Nilson's sphere decoder with S-E for 3D MIMO code and proposed simplified ML decoders, in quasi-static Rayleigh channel with 16-QAM modulation.}
\label{fig_division_16QAM}
\end{figure}

Finally, Figure~\ref{fig_division_QPSK} and Figure~\ref{fig_division_16QAM} present the overall divisions spent by the decoders.
The proposed decoders require less divisions than the Guo-Nilson's.
For instance, the proposed decoder with 2-by-2 column switch requires $47\%$, $9\%$ and $5\%$ less divisions at SNR of $0$~dB, $10$~dB and $20$~dB, respectively, with QPSK.
In 16-QAM case, it achieves $79\%$ reduction of divisions at $8$~dB.
In the meantime, the two decoders spend roughly the same number of divisions in higher SNR region, e.g. greater than $18$~dB.

In general, we can see the different trade-offs achieved by different decoders.
Proposed decoders achieves ML performance with less time latency and less divisions than Guo-Nilson's.
On the other hand, the Guo-Nilson's decoder needs less multiplications with some performance loss with 16-QAM.

\section{Conclusion}
\label{sec:conclusion}
The 3D MIMO code has been shown to be efficient and robust in distributed MIMO scenarios.
Yet, it suffers from high ML decoding complexity.
In this paper, we first proposed a new form of the 3D MIMO codeword and investigated some important properties of the new codeword. With these properties, the 3D MIMO code is proved to be fast decodable. Consequently, we proposed a reduced-complexity ML decoder for the 3D MIMO code which offers the same performance as ML decoder. Simulation results demonstrate that the novel low-complexity decoder yields much less processing time latency than the classical Guo-Nilson's sphere decoder with Schnorr-Euchner enumeration.
Moreover, the proposed 2-by-2 column switch technique can significantly reduce the average decoding complexity, especially with 16-QAM modulation.

%% ======================= algorithm =======================
%%% ------------ SimpDec --------------
\IncMargin{.5em}
\begin{algorithm}[h]
% \SetAlgoLined
 \LinesNumbered
 %\DontPrintSemicolon
 \SetKwComment{tcc}{\% }{}
 \SetKwComment{Comment}{\% }{}
 \SetCommentSty{textit}%textsl
 \SetKwFunction{QR}{QR}
 \SetKwFunction{ParaDec}{ParaDec}
 \SetKwFunction{SimpML}{SimpML}
 \SetKwFunction{enum}{S-E}
 \SetKwFunction{colswt}{ColSwt}
 \SetKwFunction{slicing}{Q}
 \SetKwInOut{Input}{$\mathbf{y}$, $\mathbf{H}$}
 \SetKwInOut{Output}{output}
\KwIn{$\widetilde{\mathbf{y}}$, $\mathbf{H}_{eq}$}
\KwOut{$\hat{\mathbf{s}}$}
 $[\mathbf{Q},\mathbf{R}]$ = \QR{$\mathbf{H}_{eq}$}\;
 $\widetilde{\mathbf{z}}=\mathbf{Q}^T\widetilde{\mathbf{y}}$, $\mathbf{s}_{_{\mathrm{ZF}}}=\mathbf{H}_{eq}^{\dag}\widetilde{\mathbf{y}}$\;
 $[\overrightarrow{\mathbf{s}}_{_{\mathrm{ZF}}}, \overrightarrow{\mathbf{H}}_{eq}]$ = \colswt($\mathbf{s}_{_{\mathrm{ZF}}}$, $\mathbf{H}_{eq}$)\label{algo_ColSwt}\Comment*[r]{\small column switch}
 ${\boldsymbol \Omega}$ = \enum{$\overrightarrow{\mathbf{s}}_{_{\mathrm{ZF}}}$}\label{algo_SE}\Comment*[r]{\small Schnorr-Euchner enumeration}
 $\mathrm{radius}=\infty$, $d=0$\;
 $\hat{\mathbf{s}}=\texttt{zeros}(16,1)$, $\overline{\mathbf{s}}=\texttt{zeros}(8,1)$\;
$l=8$\Comment*[r]{\small start from root node}
run \SimpML($\widetilde{\mathbf{z}}$, $\mathbf{R}$, $\hat{\mathbf{s}}$, $\overline{\mathbf{s}}$, $\mathrm{radius}$, $l$, $d$, $\boldsymbol \Omega$)\;

\caption{Simple ML decoder for 3D MIMO code.}\label{algo_main}
\end{algorithm}
\DecMargin{.5em}

%%% ------------ SimpDec --------------
\IncMargin{.5em}
\begin{algorithm}[h]
% \SetAlgoLined
 \LinesNumbered
 %\DontPrintSemicolon
 \SetKwComment{tcc}{\% }{}
 \SetKwComment{Comment}{\% }{}
 \SetCommentSty{textit}
 \SetKwFunction{QR}{QR}
 \SetKwFunction{ParaDec}{ParaDec}
 \SetKwFunction{SimpML}{SimpML}
 \SetKwFunction{enum}{S-E}
 \SetKwFunction{slicing}{Q}
 \SetKwInOut{Input}{$\mathbf{y}$, $\mathbf{H}$}
 \SetKwInOut{Output}{output}
 \KwIn{$\widetilde{\mathbf{z}}$, $\mathbf{R}$, $\hat{\mathbf{s}}$, $\overline{\mathbf{s}}$, $\mathrm{radius}$, $l$, $d$, $\boldsymbol \Omega$}
 \KwOut{$\hat{\mathbf{s}}$, $\mathrm{radius}$, $d$}
$d_{new}=d_{p}=d_t=0$\;
$\widetilde{\mathbf{z}}_{14}=\widetilde{\mathbf{z}}(1:8)$, $\widetilde{\mathbf{z}}_{58}=\widetilde{\mathbf{z}}(9:16)$\;
    \For(\tcc*[f]{\small check all nodes}){$j= 1$ \KwTo $\sqrt{M}$}{
        $\overline{\mathbf{s}}(l)={\boldsymbol \Omega}(l,j)$\;
        $d_{new}=|\widetilde{\mathbf{z}}_{58}(l)-\mathbf{R}{\tiny (l+8,l+ 8:16)}\overline{\mathbf{s}}(l:8)|^2 + d(l)$\Comment*[r]{\small overall distance of $[\hat{s}_5\ldots,\hat{s}_8]^T$}
        \If(\tcc*[f]{\small inside sphere}){$d_{new}<\mathrm{radius}$}{
            \eIf(\tcc*[f]{\small tree search phase}){$l\neq1$}
            {
                run \SimpML($\widetilde{\mathbf{z}}$, $\mathbf{R}$, $\hat{\mathbf{s}}$, $\overline{\mathbf{s}}$, $\mathrm{radius}$, $l-1$, $d_{new}$, $\boldsymbol \Omega$)\Comment*[r]{\small check lower layer}
            }(\tcc*[f]{\small leaf node found}){
                $\hat{\mathbf{c}}=\overline{\mathbf{s}}(1:4)$, $\hat{\mathbf{d}}=\overline{\mathbf{s}}(5:8)$\;
                compute $v_1$, $v_2$, $v_3$ and $v_4$\;
                run \ParaDec{$v_1$, $v_2$, $v_3$, $v_4$, $\mathbf{R}$, $\mathrm{radius}$, $d_{new}$}\Comment*[r]{\small parallel decision phase}
                 $d_{t}=d_p+d_{new}$\Comment*[r]{\small overall distance}
                 \If(\tcc*[f]{\small better solution found}){$d_{t}<\mathrm{radius}$\label{algo_radius_cmp}}{
                    $\hat{\mathbf{s}}=[\hat{\mathbf{a}}, \hat{\mathbf{b}}, \hat{\mathbf{c}}, \hat{\mathbf{d}}]^T$\;
                    $\mathrm{radius}=d_{t}$\label{algo_radius_new};
                 }
            }
        }
    }
 \caption{Simple ML decoder \texttt{SimpML}.}\label{algo_simp_tree}
\end{algorithm}
\DecMargin{.5em}

%% ------------------- ParaDec ----------------

\IncMargin{.5em}
\begin{algorithm}[h]
% \SetAlgoLined
 \LinesNumbered
 %\DontPrintSemicolon
 \SetKwComment{tcc}{\% }{}
 \SetKwComment{Comment}{\% }{}
 \SetCommentSty{textit}
 \SetKwFunction{QR}{QR}
 \SetKwFunction{simpML}{ParaDec}
 \SetKwFunction{enum}{S-E}
 \SetKwFunction{slicing}{Q}
 \SetKwInOut{Input}{$\mathbf{y}$, $\mathbf{H}$}
 \SetKwInOut{Output}{output}
\KwIn{$v_1$, $v_2$, $v_3$, $v_4$, $\mathbf{R}$, $\mathrm{radius}$, $d$}
\KwOut{$\hat{\mathbf{a}}$, $\hat{\mathbf{b}}$, $d_p$}
$\mathbf{R}_{11}=\mathbf{R}(1:4,1:4)$, $\mathbf{R}_{22}=\mathbf{R}(5:8,5:8)$\;
$\mathrm{flag}_1=\mathrm{flag}_2=\mathrm{flag}_3=\mathrm{flag}_4=1$\Comment*[r]{\small decision flags}
$\tau_1=\tau_2=\tau_3=\tau_4=\infty$\;
$p_1=p_2=p_3=p_4=\infty$\;%\Comment*[r]{\small current minimum distances}
$d_1=d_2=d_3=d_4=\infty$\;%\Comment*[r]{\small distances of the best solution}
  \For{$j= 1$ \KwTo $\sqrt{M}$}{
    \If{$\mathrm{flag}_1==1$\label{alg_chk_start}}{
        $\overline{s}_2^R={\boldsymbol \Omega}(3,j)$ \label{algo_s2R}\;
        $\tau_1=|v_2^R-\mathbf{R}_{11}(3,3)\overline{s}_2^R|^2$ \label{algo_tau1} \Comment*[r]{\small current distance}
    }
    \If{$\tau_1>p_{1}||(\tau_1+d_2+d_3+d_4+d)>\mathrm{radius}$}{
        $\mathrm{flag}_1=0$\Comment*[r]{\small stop search in this branch}
        $d_1=p_1$\label{alg_chk_end}\Comment*[r]{\small minimum distance of the branch}
    }
    \If{$\mathrm{flag}_1=\mathrm{flag}_2=\mathrm{flag}_3=\mathrm{flag}_4=0$}{
        \textbf{break}\Comment*[r]{\small terminate if all branches stop}
    }
    repeat lines~\ref{alg_chk_start} to~\ref{alg_chk_end} for $\overline{s}_2^I$, $\overline{s}_4^R$ and $\overline{s}_4^I$\;
    \If{$\mathrm{flag}_1==1$\label{alg_dc_start}}{
        $\overline{s}_1^R$ = \slicing{$(v_1^{R}-\mathbf{R}_{11}(1,3)\overline{s}_2^R)/(\mathbf{R}_{11}(1,1))$}\;
        $\tau_{1}^{\prime}=|v_1^R-\mathbf{R}_{11}(1,1)\overline{s}_1^R-\mathbf{R}_{11}(1,3)\overline{s}_2^R|^2+\tau_1$\Comment*[r]{\small current distance of the branch}
        \If{$\tau_{1}^{\prime}<p_1$}{
            $\hat{s}_1^R=\overline{s}_1^R$, $\hat{s}_2^R=\overline{s}^R_2$\Comment*[r]{\small current best solutions}
            $p_1=\tau_{1}^{\prime}$~\label{algo_radius_up}\Comment*[r]{\small current minimum branch distance}
        }
        \label{alg_dc_end}
    }
    repeat lines~\ref{alg_dc_start} to~\ref{alg_dc_end} for other branches\;
 }
    $\hat{\mathbf{a}}=[\hat{s}_1^R, \hat{s}_1^I, \hat{s}_2^R, \hat{s}_2^I]$,
    $\hat{\mathbf{b}}=[\hat{s}_3^R, \hat{s}_3^I, \hat{s}_4^R, \hat{s}_4^I]$\Comment*[r]{\small best solution}
    $d_p=d_1+d_2+d_3+d_4$\Comment*[r]{\small overall distance of $[\hat{s}_1,\ldots\hat{s}_4]^T$}

 \caption{Parallel decision algorithm \texttt{ParaDec}.}\label{algo_para_dec}
\end{algorithm}
\DecMargin{.5em}

\IncMargin{.5em}
\begin{algorithm}[h]
 %\SetAlgoLined
  \LinesNumbered
 %\DontPrintSemicolon
 \SetKwComment{tcc}{\% }{}
 \SetKwComment{Comment}{\% }{}
 \SetCommentSty{textit}%textsl
 \SetKwFunction{QR}{QR}
 \SetKwFunction{ParaDec}{ParaDec}
 \SetKwFunction{SimpML}{SimpML}
 \SetKwFunction{enum}{S-E}
 \SetKwFunction{slicing}{Q}
 \SetKwInOut{Input}{$\mathbf{y}$, $\mathbf{H}$}
 \SetKwInOut{Output}{output}
\KwIn{$\mathbf{s}_{_{\mathrm{ZF}}}$, $\mathbf{H}_{eq}$}
\KwOut{$\overrightarrow{\mathbf{H}}_{eq}$,$\overrightarrow{\mathbf{s}}_{_{\mathrm{ZF}}}$}
compute $\epsilon_{jk}$'s \;
    \eIf(\tcc*[f]{\small decode $[s_5,\ldots s_8]$ by tree search}){$\epsilon_{14}<\epsilon_{58}$\label{algo_cmp_2half}}{
       $\overrightarrow{\mathbf{s}}=[s_1,s_2,s_3,s_4,s_5,s_6,s_7,s_8]$ \;
        \If(\tcc*[f]{\small valid only in 2-by-2 column switch}){$\epsilon_{78}<\epsilon_{56}$}{
           $\overrightarrow{\mathbf{s}}=[s_3,s_4,s_1,s_2,s_7,s_8,s_5,s_6]$ \;
        }
    }(\tcc*[f]{\small decode $[s_1,\ldots s_4]$ by tree search})
    {
       $\overrightarrow{\mathbf{s}}=[s_5,s_6,s_7,s_8,s_1,s_2,s_3,s_4]$ \;
        \If(\tcc*[f]{\small valid only in 2-by-2 column switch}){$\epsilon_{34}<\epsilon_{12}$}{
           $\overrightarrow{\mathbf{s}}=[s_7,s_8,s_5,s_6,s_3,s_4,s_1,s_2]$ \;
        }
    }
    permute $\mathbf{s}_{_{\mathrm{ZF}}}$ and $\mathbf{H}_{eq}$ according to $\overrightarrow{\mathbf{s}}$\;
    \Return permutation results $\overrightarrow{\mathbf{H}}_{eq}$, $\overrightarrow{\mathbf{s}}_{_{\mathrm{ZF}}}$.
\caption{Column switch algorithm \texttt{ColSwt}.}\label{algo_col_swt}
\end{algorithm}

\section*{Appendices}

\subsection*{Definition of QR decomposition}
\label{sec:appdx1}
If we write $\mathbf{H}_{eq}=[\mathbf{h}_1,\ldots,\mathbf{h}_{2\kappa}]$, the
$\mathbf{H}_{eq}$'s QR decomposition $\mathbf{H}_{eq}=\mathbf{Q}\mathbf{R}$ is achieved by Gram-Schmidt procedure such that:
$\mathbf{Q}\triangleq[\mathbf{q}_1,\ldots,\mathbf{q}_{2\kappa}]$, where columns $\mathbf{q}_j$'s are orthogonal, and
\begin{equation}
   \mathbf{R}\triangleq\left [\begin{array}{*{20}c}
  \|\mathbf{r}_1\|^2 & \langle \mathbf{q}_1,\mathbf{h}_2\rangle & \cdots & \langle \mathbf{q}_1,\mathbf{h}_{2\kappa}\rangle \\
  0 & \|\mathbf{r}_2\|^2 & \cdots & \langle \mathbf{q}_2,\mathbf{h}_{2\kappa}\rangle \\
  \vdots & \vdots & \ddots & \vdots \\
  0&0 &\cdots & \|\mathbf{r}_{2\kappa}\|^2 \\
 \end{array}\right],
\end{equation}
where $\mathbf{r}_1=\mathbf{h}_1$, $\mathbf{r}_{j}=\mathbf{h}_j-\sum_{k=1}^{j-1}\langle \mathbf{q}_k,\mathbf{h}_j\rangle \mathbf{q}_k$,
$\mathbf{q}_j=\mathbf{r}_j/\|\mathbf{r}_j\|$, $j=1,\ldots,2\kappa$.

\subsection*{Proof of Theorem~\ref{theo:R11}}
\label{sec:appdx2}

Based on $\mathbf{H}_{eq}$, after some straightforward computation, it yields $\langle \mathbf{h}_1,\mathbf{h}_2\rangle=\langle \mathbf{h}_1,\mathbf{h}_4\rangle=\langle \mathbf{h}_2,\mathbf{h}_3\rangle=\langle \mathbf{h}_3,\mathbf{h}_4\rangle=0$.
According to the definition of QR decomposition, $\mathbf{q}_1=\mathbf{h}_1/\|\mathbf{h}_1\|$. Hence, $\langle \mathbf{q}_1,\mathbf{h}_2\rangle=\langle \mathbf{q}_1,\mathbf{h}_4\rangle=0$.

In addition, $\mathbf{r}_{2}=\mathbf{h}_2-\langle \mathbf{q}_1,\mathbf{h}_2\rangle \mathbf{q}_1=\mathbf{h}_2$, $\mathbf{q}_2=\mathbf{r}_2/\|\mathbf{r}_2\|= \mathbf{h}_2/\|\mathbf{h}_2\|$.
Taking into account that $\langle \mathbf{h}_2,\mathbf{h}_3\rangle=0$, it yields $\langle \mathbf{q}_2,\mathbf{h}_3\rangle=0$.

Moreover, $\mathbf{r}_{3}=\mathbf{h}_3-\sum_{j=1}^{2}\langle \mathbf{q}_j,\mathbf{h}_3\rangle \mathbf{q}_j=\mathbf{h}_3-\langle \mathbf{q}_1,\mathbf{h}_3\rangle\mathbf{q}_1$ and
$\mathbf{q}_3=\mathbf{r}_3/\|\mathbf{r}_3\|=(\mathbf{h}_3-\langle \mathbf{q}_1,\mathbf{h}_3\rangle \mathbf{q}_1)/\|\mathbf{r}_3\|$. Therefore, $\langle \mathbf{q}_3,\mathbf{h}_4\rangle=(\langle \mathbf{h}_3,\mathbf{h}_4\rangle-\langle \mathbf{q}_1,\mathbf{h}_3\rangle\langle \mathbf{q}_1,\mathbf{h}_4\rangle)/\|\mathbf{r}_3\|=0$.

This completes the proof of Theorem~\ref{theo:R11}.

\subsection*{Proof of Theorem~\ref{theo:R12}}
\label{sec:appdx3}

Based on $\mathbf{H}_{eq}$, after some straightforward computation, it yields $\langle \mathbf{h}_j,\mathbf{h}_k\rangle=0$, $\forall j=1,2,3,4$ and $k=5,6,7,8$.

Using
$\mathbf{q}_1=\mathbf{h}_1/\|\mathbf{h}_1\|$ and $\mathbf{q}_2=\mathbf{h}_2/\|\mathbf{h}_2\|$
which have been proven in the proof of Theorem~\ref{theo:R11}, it yields
$\langle \mathbf{q}_j,\mathbf{h}_k\rangle=0$, $\forall j=1,2$ and $k=5,6,7,8$.

Using
$\mathbf{q}_3=(\mathbf{h}_3-\langle \mathbf{q}_1,\mathbf{h}_3\rangle \mathbf{q}_1)/\|\mathbf{r}_3\|$ which has been proven in the proof of Theorem~\ref{theo:R11}, it yields $\langle\mathbf{q}_3, \mathbf{h}_k\rangle=(\langle \mathbf{h}_3, \mathbf{h}_k\rangle - \langle \mathbf{q}_1,\mathbf{h}_3\rangle \langle\mathbf{q}_1, \mathbf{h}_k\rangle)/\|\mathbf{r}_3\|=0$, $\forall k=5,6,7,8$.

Similarly, since $\mathbf{q}_4=(\mathbf{h}_4-\langle \mathbf{q}_2,\mathbf{h}_4\rangle \mathbf{q}_2)/\|\mathbf{r}_4\|$, it yields $\langle\mathbf{q}_4, \mathbf{h}_k\rangle=(\langle \mathbf{h}_4, \mathbf{h}_k\rangle - \langle \mathbf{q}_2,\mathbf{h}_4\rangle \langle\mathbf{q}_2, \mathbf{h}_k\rangle)/\|\mathbf{r}_4\|=0$, $\forall k=5,6,7,8$.

This completes the proof of Theorem~\ref{theo:R12}.

\subsection*{Proof of Corollary~\ref{cor:R22}}
\label{sec:appdx4}

Using the similar method as in the proof of Theorem~\ref{theo:R11}, it can be computed from the definition of $\mathbf{H}_{eq}$ that
$\langle \mathbf{h}_5,\mathbf{h}_6\rangle=\langle \mathbf{h}_5,\mathbf{h}_8\rangle=\langle \mathbf{h}_6,\mathbf{h}_7\rangle=\langle \mathbf{h}_7,\mathbf{h}_8\rangle=0$.

In addition, using Theorem~\ref{theo:R12}, it can be obtained that $\mathbf{q}_5=\mathbf{h}_5/\|\mathbf{h}_5\|$. Hence,
$\langle \mathbf{q}_5,\mathbf{h}_6\rangle=\langle \mathbf{q}_5,\mathbf{h}_8\rangle=0$.

Using $\langle \mathbf{q}_5,\mathbf{h}_6\rangle=0$ and Theorem~\ref{theo:R12}, it yields $\mathbf{q}_{6}=\mathbf{h}_6/\|\mathbf{h}_6\|$.
Hence, $\langle \mathbf{q}_6,\mathbf{h}_7\rangle=0$.

Finally, using $\langle \mathbf{q}_6,\mathbf{h}_7\rangle=0$ and Theorem~\ref{theo:R12}, it yields
$\mathbf{r}_{7}=\mathbf{h}_7-\langle \mathbf{q}_5,\mathbf{h}_7\rangle\mathbf{q}_5$
and
$\mathbf{q}_7=\mathbf{r}_7/\|\mathbf{r}_7\|=(\mathbf{h}_7-\langle \mathbf{q}_5,\mathbf{h}_7\rangle \mathbf{q}_5)/\|\mathbf{r}_7\|$.
Therefore, $\langle \mathbf{q}_7,\mathbf{h}_8\rangle=(\langle\mathbf{h}_7, \mathbf{h}_8\rangle-\langle \mathbf{q}_5,\mathbf{h}_7\rangle \langle \mathbf{q}_5, \mathbf{h}_8\rangle)/\|\mathbf{r}_7\|=0$.

This completes the proof of Corollary~\ref{cor:R22}.

%%%%%%%%%%%%%%%%%%%%%%%%%%%
\section*{Acknowledgements}
  \ifthenelse{\boolean{publ}}{\small}{}
This work is supported by French ANR  ``Mobile Multi-Media (M3)'' project and ``P\^ole Images \& R\'eseaux''.

%%  Bmc_article.bst  will be used to                       %%
%%  create a .BBL file for submission, which includes      %%
%%  XML structured for BMC.                                %%
%%  After submission of the .TEX file,                     %%
%%  you will be prompted to submit your .BBL file.         %%

\newpage
{\ifthenelse{\boolean{publ}}{\footnotesize}{\small}
 \bibliographystyle{bmc_article}  % Style BST file
  \bibliography{bmc_article} }     % Bibliography file (usually '*.bib' )

%%%%%%%%%%%

\ifthenelse{\boolean{publ}}{\end{multicols}}{}

%%%%%%%%%%%%%%%%%%%%%%%%%%%%%%%%%%%
%%                               %%
%% Figures                       %%
%%                               %%
%% NB: this is for captions and  %%
%% Titles. All graphics must be  %%
%% submitted separately and NOT  %%
%% included in the Tex document  %%
%%                               %%
%%%%%%%%%%%%%%%%%%%%%%%%%%%%%%%%%%%

%%
%% Do not use \listoffigures as most will included as separate files

\section*{Figures}
  \subsection*{Figure 1 - $\mathbf{R}$ matrix of the new 3D MIMO codeword}
      Illustration of the $\mathbf{R}$ matrix of the new 3D MIMO codeword given in (\ref{eq:3Dnew}) in quasi-static channel.

  \subsection*{Figure 2 - $\textbf{R}$ matrix of the original 3D MIMO codeword}
      Illustration of the $\textbf{R}$ matrix of the original 3D MIMO codeword given in (\ref{eq:3D}) in quasi-static channel.
  \subsection*{Figure 3 - Two-stage sphere decoding}
      Illustration of the two-stage sphere decoding.

  \subsection*{Figure 4 - Parallel decisions}
      Parallel decisions of $[s_1^R,s_2^R]$, $[s_1^I,s_2^I]$, $[s_3^R,s_4^R]$ and $[s_3^I,s_4^I]$.

  \subsection*{Figure 5 - Symbol error rate comparison}
       Symbol error rate obtained with Guo-Nilson's sphere decoder with S-E for 3D MIMO code, ML decoder and proposed simplified ML decoders in quasi-static Rayleigh channel with QPSK and 16QAM.

  \subsection*{Figure 6 - Complexity in terms of visited nodes with QPSK}
      Computational complexity in terms of visited nodes required by Guo-Nilson's sphere decoder with S-E for 3D MIMO code and proposed simplified ML decoders, in quasi-static Rayleigh channel with QPSK modulation.

  \subsection*{Figure 7 - Complexity in terms of visited nodes with 16-QAM}
      Computational complexity in terms of visited nodes required by Guo-Nilson's sphere decoder with S-E for 3D MIMO code and proposed simplified ML decoders, in quasi-static Rayleigh channel with 16-QAM modulation.

  \subsection*{Figure 8 - Complexity in terms of multiplications with QPSK}
      Computational complexity in terms of multiplications required by Guo-Nilson's sphere decoder with S-E for 3D MIMO code and proposed simplified ML decoders, in quasi-static Rayleigh channel with QPSK modulation.

  \subsection*{Figure 9 - Complexity in terms of multiplications with 16-QAM}
      Computational complexity in terms of multiplications required by Guo-Nilson's sphere decoder with S-E for 3D MIMO code and proposed simplified ML decoders, in quasi-static Rayleigh channel with 16-QAM modulation.

  \subsection*{Figure 10 - Complexity in terms of divisions with QPSK}
      Computational complexity in terms of divisions required by Guo-Nilson's sphere decoder with S-E for 3D MIMO code and proposed simplified ML decoders, in quasi-static Rayleigh channel with QPSK modulation.

  \subsection*{Figure 11 - Complexity in terms of divisions with 16-QAM}
      Computational complexity in terms of divisions required by Guo-Nilson's sphere decoder with S-E for 3D MIMO code and proposed simplified ML decoders, in quasi-static Rayleigh channel with 16-QAM modulation.

%%%%%%%%%%%%%%%%%%%%%%%%%%%%%%%%%%%
%%                               %%
%% Tables                        %%
%%                               %%
%%%%%%%%%%%%%%%%%%%%%%%%%%%%%%%%%%%

%% Use of \listoftables is discouraged.
%%
\section*{Tables}
  \subsection*{Table 1 - Comparison of ML decoding complexities of STBCs for $4\times2$ MIMO transmission}
    The table presents the ML decoding complexities of several STBCs that are suitable for $4\times2$ MIMO transmission.\par \mbox{}
%    \par
%    \mbox{
%
%      }

%%%%%%%%%%%%%%%%%%%%%%%%%%%%%%%%%%%
%%                               %%
%% Additional Files              %%
%%                               %%
%%%%%%%%%%%%%%%%%%%%%%%%%%%%%%%%%%%

%\section*{Additional Files}
%  \subsection*{Additional file 1 --- Sample additional file title}
%    Additional file descriptions text (including details of how to
%    view the file, if it is in a non-standard format or the file extension).  This might
%    refer to a multi-page table or a figure.
%
%  \subsection*{Additional file 2 --- Sample additional file title}
%    Additional file descriptions text.

\end{bmcformat}
\end{document}